%% file: paper.tex
\documentclass[a4paper,UKenglish,cleveref, autoref, thm-restate]{lipics-v2021}

\newcommand{\abs}[1]{{\left | #1 \right |}}

\renewcommand{\paragraph}[1]{\medskip \noindent {\bf #1.}}

\newcommand{\M}{\mathcal{M}}

\newcommand{\ANN}{\mathtt{ANN}}

\newcommand{\eps}{\epsilon}
\newcommand{\veps}{\varepsilon}

\DeclareMathOperator{\E}{{\mathop{\mathbf{E}}}}

\def\<{\langle}
\def\>{\rangle}
\def\RR{\mathbb{R}}
\def\CC{\mathbb{C}}

\def\EE{\mathbf{E}}
\def\PP{\mathbf{Pr}}

\usepackage{physics}

\newcommand{\rev}[1]{{\color{black}#1}}
\newcommand{\revall}[1]{{\color{black}#1}}
\newcommand{\revone}[1]{{\color{black}#1}}

\newcommand{\revfive}[1]{{\color{black}#1}}

\usepackage[linesnumbered,ruled,vlined,procnumbered]{algorithm2e}
\usepackage{algpseudocode}

\nolinenumbers

\title{Quantum Data Sketches} 

\author{Qin Zhang}{Indiana University Bloomington, USA \and \url{https://homes.luddy.indiana.edu/qzhangcs/} }{qzhangcs@iu.edu}{https://orcid.org/0000-0002-6851-3115}{supported by NSF CCF-1844234 and IU Luddy Faculty Fellowship}

\author{Mohsen Heidari}{Indiana University Bloomington, USA \and \url{https://homes.luddy.indiana.edu/mheidar/}}{mheidar@iu.edu }{}{supported by NSF CCF-2211423}

\authorrunning{Q. Zhang and M. Heidari} 

\Copyright{Qin Zhang and Mohsen Heidari} 

\ccsdesc[500]{Theory of computation~Quantum query complexity}
\ccsdesc[500]{Information systems~Query operators}
\ccsdesc[500]{Information systems~Query optimization}

\keywords{quantum data representation, data sketching, query execution} 

\EventEditors{John Q. Open and Joan R. Access}
\EventNoEds{2}
\EventLongTitle{42nd Conference on Very Important Topics (CVIT 2016)}
\EventShortTitle{CVIT 2016}
\EventAcronym{CVIT}
\EventYear{2016}
\EventDate{December 24--27, 2016}
\EventLocation{Little Whinging, United Kingdom}
\EventLogo{}
\SeriesVolume{42}
\ArticleNo{23}

\begin{document}

\maketitle

\begin{abstract}
Recent advancements in quantum technologies, particularly in quantum sensing and simulation, have facilitated the generation and analysis of inherently quantum data. This progress underscores the necessity for developing efficient and scalable quantum data management strategies. This goal faces immense challenges due to the exponential dimensionality of quantum data and its unique quantum properties such as no-cloning and measurement stochasticity. Specifically, classical storage and manipulation of an arbitrary $n$-qubit quantum state requires exponential space and time. Hence, there is a critical need to revisit foundational data management concepts and algorithms for quantum data.   In this paper, we propose succinct quantum data sketches to support basic database operations such as search and selection.  We view our work as an initial step towards the development of quantum data management model, opening up many possibilities for future research in this direction.
\end{abstract}

\input{intro}

\input{related}

\input{preliminary}

\input{query}

\input{data}

\input{future}



\bibliography{paper}

\appendix
\input{appendix}

\end{document}

%% file: intro.tex
\section{Introduction}
\label{sec:intro}

Quantum information and computing, rooted in the principles of quantum mechanics, have emerged as an important field of study with far-reaching effects across a broad spectrum of disciplines. Central to the concept of quantum computing are quantum bits (or qubits), which set themselves apart from classical bits due to their ability to exist in a superposition of states, allowing a quantum computer to offer the potential computational advantage against classical computing.

Although significant advancements have been made in the development of quantum algorithms after several decades of research, only a handful provably outperform their classical counterparts.  Notable examples include
Shor's algorithm for factorization~\cite{Shor97}, Grover's algorithm for search~\cite{Grover96}, and linear system solvers~\cite{HHL09}.  These quantum algorithms typically start by encoding classical input data into quantum states, execute a series of quantum operations, and then measure the resulting quantum states and carry out specific post-processing on the measurement outcomes. 
The reasons for the difficulties in the design of quantum algorithms that can outperform classical counterparts on {\em classical input data}  remain elusive.  

In this paper, we take a different perspective, directing our attention towards quantum data themselves. The nature, along with scientific experiments spanning physics, chemistry, material science, biology, and other fields, generates massive quantities of quantum data every day.  Sources include Hawking Radiation, Cosmic Microwave Background, quantum effects in neutron stars, quantum states in ultra-cold atoms, quantum information in DNA replication, etc.  
\revall{
In many scenarios, there is a need for us to preserve quantum data that has been collected from nature or generated in labs for future analysis. For example, scientists often use photons collected from remote stars to study the properties of those astronomical objects. It would be beneficial to store those photons as quantum states in a database, since it may not be feasible to collect fresh photons from those astronomical objects at the time of data analysis.  In the case that the quantum states are prepared in the labs, generating fresh copies of quantum states on demand is often time-consuming. Let us use quantum simulation as an example. 
Quantum simulation is a prominent advantage of quantum computers, with significant implications for numerous areas of scientific research, including computer-aided drug design~\cite{PAB+23}, high-energy physics \cite{Jordan2012}, quantum chemistry \cite{Whitfield2011} and many-body physics \cite{SKPK19}.  Quantum simulation typically relies on solving the Schr\"odinger equation for the underlying Hamiltonian. The Hamiltonian is implemented by a quantum circuit, which is applied to an initial quantum state to generate target quantum states.  
The construction of the Hamiltonians and the preparation of the target states can be rather time-consuming.\footnote{For instance, the Hamiltonian of the two-dimensional Fermi-Hubbard model on an $8 \times 8$ lattice already requires approximately $10^7$ Toffoli gates \cite{Kivlichan2020}, which directly contribute to the query time if states need to be generated from scratch at query.} Storing the generated molecular quantum states in a database would eliminate the need to repeat the state preparation procedures during data analysis.
}

\revall{Once the quantum states are stored in a database, and assuming each state is associated with additional information such as the nature sources recorded at the time of collection or parameters of the experimental setup used to produce them, numerous applications can be envisioned.
For example, if scientists receive photons from an unknown remote star, they can search a photon database to find a matching quantum state. Upon finding a match, they can retrieve its associated properties and other information, such as the time and method of its previous observation.  They may also want to sort the states using a local observable (see Definition~\ref{def:local-ob} in Section~\ref{sec:query}) with respect to certain properties (such as energy or momentum) to get an order of the photons in the database, aiding in the understanding of the spectrum of the corresponding stars in the universe. In quantum simulation, if we want to produce molecular states with average energy levels above a certain threshold relative to a specific local observable, we can perform a selection operation in our database to identify those states, and then use the associated parameters for the experimental setup to produce more of such quantum states.}

Nevertheless, quantum data management remains in its infant stage.  Some of the previously mentioned motivating examples are more like anticipated future problems. 
There has been research that leverages quantum data for learning or optimization, such as quantum machine learning~\cite{Heidari2021,Haah2017,Aaronson2004}, quantum variational optimization algorithm~\cite{Harrow2021,Farhi2014}. and quantum neural network~\cite{Schuld2014,Mitarai2018,Farhi2018,HeidariAAAI2022,Massoli2021,Garg2020}. However, their primary focus is on the sample complexity (namely, the number of copies of the quantum state needed for the task) and the convergence to optimal points, rather than on developing methods for the efficient representation and storage of quantum data for subsequent analysis.

In this paper, we introduce several quantum data sketches to support basic database operations in a {\em sustainable} and {\em efficient} manner. This paper does not aim to formulate a comprehensive quantum data management model. Rather, we view our work as an initial step towards developing a sustainable model for representing, querying, and analyzing quantum data at scale.

\paragraph{Unique Challenges in the Quantum World}
The quantum world possesses several unique properties, such as superposition and entanglement, that can be leveraged to reduce resource usage in computing and information exchange. However, some of these features also post significant challenges to quantum data management. We highlight a few below.

{\em Post-Measurement State Disturbance.} \ The only way to extract information from a quantum state is to perform quantum measurements and observe probabilistic outcomes. However, each measurement has the effect of perturbing the quantum state. 
This characteristic implies that a quantum state might not be reusable post-measurement.   In other words, we may need to consume many identical copies of a quantum state in order to derive enough useful information about it.  This phenomenon is in stark contrast with the classical setting, in which we can consistently access the same data element for a number of times, always yielding the same result.  

{\em No-cloning.} \ A natural thought to resolve the issue caused by state disturbance is to clone the quantum state before the operations.  Unfortunately, the {\em no-cloning theorem} (see, e.g., \cite{NC10}) in quantum mechanics asserts that it is impossible to create an exact copy of an arbitrary unknown quantum state.  

{\em Lack of Large-Scale Quantum Storage Systems.}  \ 
At the time of writing this paper, we are not aware of any reliable large-scale quantum storage systems. One reason for this is that qubits are highly susceptible to environmental disruptions such as temperature variations, electromagnetic radiation, or particle interactions. These disruptions lead to what is known as decoherence~\cite{JZK+03}, resulting in the loss of quantum information. 

Moreover, due to the quantum state disturbance and the no-cloning principle, even if we successfully build viable large-scale quantum storage systems in the future, we still need many identical copies of the quantum state for any nontrivial database operation. 
This implies that in order to accommodate an unlimited number of database operations (i.e., to be sustainable), we must prepare an unlimited number of copies for each quantum state in the storage, which is certainly {\em not} practical.  

An alternative approach is to first learn the classical description of each quantum state and store it in a classical memory for future operations.  Indeed, we believe that for the purpose of quantum data management, we have to store quantum states in the classical format.  However, learning and storing the full information of a quantum state as a classical object is both time and space expensive, as the dimensionality of a quantum state is exponential in terms of the number of qubits.

We thus propose to design {\em succinct classical representations} (or, sketches) of quantum states that can be used to perform database operations efficiently.  Based on the particular database operation it is intended to support, each sketch preserves only {\em partial} information of a quantum state.  This is also the reason why we may be able to make the size of the sketch to be $o(d)$, where $d$ is the dimension of the quantum state.  We also note that the sample complexity for constructing data sketches is a secondary consideration for database management systems, as it is just a one-time preprocessing step in the database design. This is where our work departs from the quantum state learning/tomography literature, which we will discuss in Section~\ref{sec:related}.

\paragraph{Our Contribution}
We give the first systematic approach to designing {\em space-efficient} sketches for quantum states. These sketches can then be used to develop {\em time-efficient} algorithms for basic database operations.  In particular:
\begin{enumerate}
    \item In Section~\ref{sec:query}, we have formalized a set of basic database operations for quantum data, including search, selection, sorting, and join. These operations differ from those for classical data as they inherently incorporate approximation in their definitions.
    
    \item Our main technical results are the first set of classical vector sketches that preserve, up to a distortion of $(1 + \iota)$ for an arbitrarily small $\iota > 0$, the trace distance of the quantum states with probability $(1 - \delta)$. Our sketches have sizes $O(\log(1/\delta)/\iota^2)$, which is {\em independent} of the dimension of the states.  Coupled with efficient nearest neighbor search via locality sensitive hashing, they can be used to support the search and join operations with time sublinear in the database size and independent of the state dimension.  See Section~\ref{sec:vector}.

    \item  We make use of classical shadow seeds of quantum states~\cite{HKP20} to approximate the expectation value of any given $k$-local observable (to be defined in Section~\ref{sec:selection}) using time and space {\em independent} of the dimension of the state. We also present a new hybrid quantum-classical algorithm to accelerate the query time. This sketch can be used for selection and sorting operations.  See Section~\ref{sec:shadow}.
\end{enumerate}

\paragraph{Paper Outline}  In Section~\ref{sec:pre}, we review some background on quantum information and computing as well as tools for classical data management.  In Section~\ref{sec:query}, we define a set of basic database operations for quantum data.  After these preparations, in Section~\ref{sec:sketch}, we present our classical sketches of quantum states and illustrate how to perform various database operations using these sketches.  We review works that are most relevant to this paper in Section~\ref{sec:related} and propose several directions for future research in Section~\ref{sec:future}.

%% file: related.tex
\subsection{Related Work}
\label{sec:related}

\revfive{We are not aware of any prior work on designing classical sketches of quantum data, except for the paper \cite{HKP20} discussed in Section~\ref{sec:shadow}.  There have been effort aiming to introduce quantum computing, quantum algorithms and quantum machine learning to the database community~\cite{Cockshott97,Younes07,Liu07,TK16,CGG+23,WGU+23}. We refer the readers to the recent tutorial~\cite{HHF24} for an overview of these works. However, these initiatives either attempt to design and perform database operations directly on quantum data (i.e., assuming database elements are stored as quantum states) or focused on speeding up databases query optimization and transactions on classical data, setting them apart from the objectives pursued in this paper.}

\revfive{There are works \cite{WDD+19,HPW24} focusing on applying classical data compression techniques (such as quantization) to the quantum state vector during quantum simulation. We note that our approach with sketches is quite different, as we aim to extract relevant information (often independent of the quantum states' dimension) for various database operations.}

\paragraph{Quantum State Learning}  
Many studies have explored the task of characterizing and learning properties of a quantum state using multiple copies of the state, including {\em approximate state discrimination}~\cite{CL21}, {\em quantum state discrimination}~\cite{HW12}, {\em quantum state tomography}~\cite{HHJ+16,OW16}, {\em quantum state property testing}~\cite{HLM17}, {\em quantum state certification}~\cite{BOW17}, {\em shadow tomography}~\cite{Aaronson18,BO21}, and {\em pretty good tomography}~\cite{Aaronson07}.  

In the problem of approximate state discrimination, we are promised that a query quantum state ${\phi}$ belongs to a set $S$ of quantum states. The algorithm's task is to return a state ${\psi} \in S$ such that $D({\phi}, {\psi}) \le \eps$.  The algorithm for approximate state discrimination proposed in \cite{CL21} can be used together with the equality testing to handle the search operation when the available number of copies of the query state is limited, at the cost of larger time and space complexities. However, the need of fresh copies of database states for equality testing would undermine the long-term sustainability of the database system.

The problem of quantum state discrimination is very similar: We are again promised that the query state ${\phi}$ belongs to a set $S$, but now the algorithm needs to return the {\em exact} ${\phi}$.  Harrow and Winter~\cite{HW12} gave an algorithm for this problem where the sample complexity of the query state depends on a parameter $F$, which is the maximum pairwise fidelity of states in the set $S$.

In the quantum state tomography, we wanted to learn an unknown quantum state up to a trace distance $\eps$.  Optimal sample complexity $\tilde{\Theta}(d/\eps^2)$ has been identified~\cite{HHJ+16,OW16}.

Quantum state property testing~\cite{HLM17} and quantum state certification~\cite{BOW17} can be seen as relaxations of aforementioned problems.  In the former, we are given a query state ${\phi}$ and a set $S$ of quantum states, and asked to test whether ${\phi} \in S$ or ${\phi}$ is $\eps$-far from $S$ (that is, for any state ${\psi} \in S$, we have $D({\phi}, {\psi}) > \eps$), and in the latter, we are given a query state ${\phi}$ and a known state ${\psi}$, and asked to test whether ${\phi} = {\psi}$ or $D({\phi}, {\psi}) > \eps$. 
The main issue with property testing and certification in the setting of data management is that the decision can be {\em arbitrary} even if the query state is very close to (but not the same as) a database state.

Both shadow tomography and pretty good tomography focus on approximating $\phi^\dagger M_i \phi$ for a query state ${\phi}$ and a set of {\em known} binary measurements $\{M_i\}$~\cite{Aaronson18, BO21}, or a distribution on them~\cite{Aaronson07}.  
However, these algorithms cannot be used for the $(\eta, \veps)$-selection for an arbitrary observable $M$ given at the time of query.  Their running time is also polynomial in terms of the state dimension $d$.
Recently, Gong and Aaronson~\cite{GA22} generalized shadow tomography to a {\em fixed} set of measurements with multiple outcomes.

To the best of our knowledge, all the previous work on quantum state learning focuses on the sample complexity, but {\em not} on the space complexity for representing the quantum states for various data management operations.

%% file: preliminary.tex
\section{Preliminaries}
\label{sec:pre}

We start by giving a gentle introduction of the basics of quantum information and computing, particularly for readers who are not  in the field yet.  For a comprehensive treatment on this topic, we refer the readers to standard textbooks in the field, such as \cite{NC10}. 

\paragraph{Quantum States and Qubits}
The first axiom of quantum mechanics is concerned with \textit{quantum state} as a way to describe a quantum system, such as a qubit. For accessibility of the paper we focus on pure state that are represented by complex-valued vectors. Moreover, we assume that each quantum data point is stored in $n$-qubits. Therefore, the dimentionality of the space is $d=2^n$.  In that case, the quantum stats are unit-norm vectors in $\mathbb{C}^d$.  Following the Dirac bra-ket notation, a vector $u\in \CC^d$ is simply denoted by the ket $\ket{u}$. As an example, a {\em qubit} is a $2$-dimensional vector represented as $\ket{\phi} = \alpha_0 \ket{0} + \alpha_1 \ket{1}$, where $\abs{\alpha_0}^2 + \abs{\alpha_1}^2 = 1$. This decomposition is typically called a \textit{superposition}. A well-known superposition is the state $\frac{1}{\sqrt{2}}(\ket{0}+\ket{1})$. Similarly, an $n$-qubit state is represented by a superposition as
$
\ket{u} = \sum_{x_1\cdots x_n\in \{0,1\}^n } \alpha_{x_1\cdots x_n} \ket{x_1\cdots x_n},
$
where $\sum_{x_1\cdots x_n\in \{0,1\}^n } \alpha^2_{x_1\cdots x_n}=1$.  For compactness, we use $\ket{i}$ to represent each $\ket{x_1\cdots x_n}$, where $i$ is the decimal representation of the binary string $x_1\cdots x_n$.

\paragraph{Quantum Operations} The second axiom of quantum mechanics states that the evolution of quantum states are described via unitary transformation.  A unitary transformation is represented by a unitary matrix $U$ such that $U^\dagger U=U U^\dagger =I$.  If the initial state is $\ket{\phi}$, then the evolved state is $U\ket{\phi}$. In quantum computing $U$ is typically implemented in terms of elementary quantum logical gates. In this perspective, one can study the gate complexity of implementing $U$. This axiom implies a unique feature of quantum, known as the {\em no-cloning} principle that prohibits making copies of quantum data.  As a result one needs to adopt data management procedures that abide this rule. 

\paragraph{Quantum Measurements}  The third axiom of quantum mechanics asserts that any classical information about a quantum state is obtained via {\em measuring} it.  The act of measuring a quantum system will collapse the quantum state inevitably. The specific outcome of a measurement is probabilistic and is governed by the Born's law.  These probabilities are determined by the initial state of the system and the nature of the interaction between the system and the measuring device. Measuring in an $n$-qubit system is typically modeled in the so-called computational basis.  When the quantum state is in the superposition $\ket{\phi} = \sum_i \alpha_i \ket{i}$, the outcome of the measurement in the computational basis is going to be $i\in [2^n]$ with probability $p_i = |\alpha_i|^2$. For instance, measuring the state $\frac{1}{\sqrt{2}}(\ket{0}+\ket{1})$ produces a random uniform binary output. The stochasticity of quantum measurements is another  feature that calls for probabilistic data management frameworks. Moreover, the state collapse phenomenon significantly complicates the tasks, as the quantum state cannot be entirely ``recycled'' following a measurement.  

\rev{One may attempt to think of a quantum state $\ket{\phi} = \sum_i \alpha_i \ket{i}$ - as far as
measurement is concerned - as a discrete
probability distribution $\{p_1, \ldots, p_d\}$, but there are two fundamental differences. First, the coefficients (called {\em amplitudes}) $\alpha_i$'s are complex numbers that make superposition and interference possible. Second, the probability of an outcome in quantum mechanics is found by taking the {\em absolute square} of the amplitude, that is, $p_i = |\alpha_i|^2$. }

In general, a certain measurement $\M$ on a quantum state can be obtained in three stages: (i) applying an appropriate quantum operator $U$ to the state, (ii) measuring the evolved state $U\ket{\phi}$ in the computational basis; and (iii) applying classical post processing on the measurement outcomes.  This procedure is compactly modeled as a matrix $M$ called an \textit{observable} that is multiplied by the original state $\ket{\phi}$. The eigenvalues of $M$ represent the possible values of the measurement outcomes.  Moreover, by $\M(\ket{\phi})$ we denote  the probability distribution of the measurement outcomes after applying $\M$ on $\ket{\phi}$. Because the outcomes are probabilistic, we are often interested in their expectation values.   The expectation of the outcome distributed by $\M(\ket{\phi})$ is equal to $\expval{M}{\phi}$, where $\bra{\phi}$ is the complex conjugate transpose of the vector $\ket{\phi}$.

\paragraph{Standard Math Notations Versus Dirac Notations}
As this paper is intended for an audience within the database community, we recognize that the Dirac bra-ket notation might appear unfamiliar to database researchers without a background in quantum information and computing. To simplify, in the main text we express a pure quantum state as a column vector with dimensions denoted as $d$, and use $\phi$ and $\phi^\dagger$ to denote $\ket{\phi}$ and $\bra{\phi}$, respectively.  
We use $\phi^\dagger M \phi$ to denote the expectation value $\ev{M}{\phi}$ of an observable $M$.  Throughout the paper, we reserve the notations $\phi$ and $\psi$ for quantum states.

We have also included a more formal (but still gentle) introduction of quantum information and computing using Dirac bra-ket notations in Appendix~\ref{sec:pre-full}.  We will use Dirac bra-ket notations in all the proofs in the appendix. 

\paragraph{Trace Distance}
Given two quantum states $\phi$ and $\psi$, we define their trace distance to be 
$D(\phi, \psi) = \sqrt{1 - \abs{\psi^\dagger \phi}^2}$.
The trace distance is the most widely used distance measure for quantum states in the literature.

\subsection{Performance Metrics}
In the context of quantum data, similar to classical database design, the efficiency of space and time is crucial during database initialization, indexing, and querying.  Minimizing the number of quantum state copies used for constructing sketches is also important, as obtaining state copies can be costly and they cannot be fully recycled due to post-measurement disturbance. However, as we mentioned earlier, sample complexity is a secondary consideration in the data management setting, since the sketch-building/initialization is a one-time process.

A unit-time quantum operation comprises standard single-qubit gates like the Hadamard gate, Pauli gates, phase gate, and $T$ gate, as well as a two-qubit gate, such as the Controlled-NOT (CNOT) gate, that enables entangling operations.\footnote{We refer the readers to \cite{NC10} for a detailed introduction of these gates.}  The combination of these gates is sufficient to approximate any unitary operation to arbitrary accuracy.  We call these gates {\em unit gates}, and define the {\em size} of a circuit (for representing a unitary operation) to be the number of unit gates in the circuit.

As mentioned, a typical quantum measurement $\M$ on $n$ qubit systems consists of a unitary operator $U_{\M}$ followed by measurement in computational basis and classical post processing. Assuming that the classical post processing is polynomial, the overall time cost is typically dominated by the gate complexity of $U_{\M}$.    It has been shown in \cite{STY+23} that a circuit depth of $\Theta(2^n/n)$ (i.e., $\Theta(d/\log d)$) is needed for constructing an arbitrary unitary operator $U$.   To simplify matters, we assume that both executing an arbitrary $d$-dimensional quantum measurement and preparing an arbitrary $d$-dimensional state require $O(d)$ quantum time.

\subsection{Nearest Neighbor in High Dimensions} 
\label{sec:ANN}

As quantum states are inherently high dimensional, even after effective sketching and summarization that we will illustrate in the subsequent sections, we will thus use {\em Approximate Nearest Neighbor (ANN)} via {\em Locality Sensitive Hashing (LSH)} to further speed up some database operations.  This subsection will take a brief detour from our discussion of quantum data management.

\begin{definition}[$(r, \beta)$-ANN-search]
	Let $X$ be a database containing a set of vectors in $\mathbb{R}^d$ and $q \in \mathbb{R}^d$ be a query vector. Let $dist(\cdot, \cdot)$ be a distance function. If there is at least one vector $p \in X$ with $dist(p, q) \le r$, return any $p' \in X$ with $dist(p', q) \le \beta r$. Otherwise, either return a $p' \in X$ with $dist(p', q) \le \beta r$ or return $\emptyset$.
\end{definition}

Let us focus on the case that the distance function $dist(\cdot, \cdot)$ is $\ell_1$ or $\ell_2$.  Indyk and Motwani~\cite{IM98} showed that $(r, \beta)$-ANN can be solved efficiently via LSH.  The idea is that we first apply multiple hash functions to each vector in $X$; this part can be pre-computed and stored as an indexing. At the time of query, we apply the same set of hash functions to the query vector $q$. We then run over all vectors $p \in X$ such that $p$ and $q$ collide (i.e., fall into the same bin) on at least one hash function, and return the first vector $p$ if $dist(p, q) \le \beta r$.  If no such $p$ found after traversing a certain number of vectors in $X$, we return $\emptyset$.

We will use $\ANN(q, X, r, \beta)$ to denote the $(r, \beta)$-ANN search for a query vector $q$ in database $X$.  The following is a summary of results on LSH-based ANN for $\ell_1/\ell_2$ distances.

\begin{theorem}[\cite{IM98,DII+04,AI06}]
	\label{thm:LSH}
	For $dist(\cdot, \cdot)$ being $\ell_1$ or $\ell_2$, a database $X$ of $m$ vectors, and a $d$-dimensional vector $q$, there is an algorithm that solves $\ANN(q, X, r, \beta)$ using  $O(dm + m^{1+\gamma})$ space and $O(d m^\gamma)$ classical time, where $\gamma \approx 1/\beta$ for $\ell_1$ distance and $\gamma \approx 1/\beta^2$ for $\ell_2$ distance.  
\end{theorem}

\begin{remark}
	\label{rem:LSH-search-all}
	We note that if we do not terminate the algorithm after encounter the first $p \in X$ such that $dist(p, q) \le \beta r$, then the same algorithm can return a subset $Y \subseteq X$ including {\em all} vectors $p$ such that $dist(p, q) \le r$, and excluding all vectors $p$ such that $dist(p, q) \ge \beta r$. 
\end{remark}

\begin{remark}
	\label{rem:LSH-join}
	We can also use LSH to find a set $J$ of pairs of vectors such that $J$ includes all pairs $(p, q)$ such that $dist(p, q) \le r$, and excludes all pairs $(p, q)$ such that $dist(p, q) \ge \beta r$.  To this end, we first hash all vectors, and then check the distances of all pairs of vectors that collide on at least one hash function.
\end{remark}

%% file: query.tex
\section{Basic Operations on Quantum Data}
\label{sec:query}

The characteristics of quantum information dictate that we can only obtain an {\em approximation} of a quantum state ${\phi}$ with a finite number of quantum state copies. A celebrated result in quantum state tomography states that to learn an unknown $n$-qubit quantum state ${\phi}$ up to a trace distance $\eps$, we already need $\Omega\left({d}/{\eps^2}\right)$ copies of the quantum state, where $d = 2^n$ is the dimension of ${\phi}$~\cite{HHJ+16,OW16}.  We thus consider two quantum states ${\phi}, {\psi}$ with $D({\phi}, {\psi}) \le \veps$ the same state. Consequently, all the operations that we support in a quantum database also need to be approximate. The precise definition of `approximation' varies for different operations.  

In this section, we formulate basic quantum data operations that we aim to support using our proposed sketches. When we say the return of a quantum state $\phi$, we are referring to its identifier. 

\subsection{Equality Test}
In the classical data setting, the equality test on two data objects returns $1$ if $p = q$, and returns $0$ otherwise. In the quantum setting, since we cannot distinguish two quantum states using $o(d/\veps^2)$ copies of the states if their trace distance is at most $\veps$, we need to introduce the approximation version of the equality test:

\begin{definition}[$(\veps, \beta)$-equality-test]
	Given two quantum states ${\phi}$ and ${\psi}$, output $1$ if $D({\phi}, {\psi}) \le \veps$, and $0$ if  $D({\phi}, {\psi}) > \beta \veps$.  The output can be arbitrary if $\veps < D({\phi}, {\psi}) \le \beta \veps$.
\end{definition}

In words, we consider two quantum states the same if their trace distance is at most $\veps$, and different if their trace distance is more than $\beta\veps$.  If the distance falls between the two values, then the decision can be arbitrary.  The gap between {\em yes} and {\em no} is inevitable for quantum data.  

Given two quantum states ${\phi}$ and ${\psi}$, which may be unknown, the standard method for estimating their trace distance is the swap test~\cite{BCW01}. The algorithm uses a controlled-SWAP gate (can be implemented using $O(n) = O(\log d)$ unit gates) and two single-qubit Hadamard gates. The test outputs $1$ with probability 
$\frac{1+\abs{\phi^\dagger\psi}^2}{2} = 1 - \frac{D({\phi}, {\psi})^2}{2}$,
and $0$ otherwise. Therefore, using $O_\beta\left(\frac{1}{\veps^2}\log{\frac{1}{\delta}}\right)$ such tests (the constant hidden in the big-$O$ depends on the constant $\beta$), we can differentiate the case $D({\phi}, {\psi}) > \beta \veps$ from $D({\phi}, {\psi}) \le \veps$ with a probability  $1 - \delta$.

The main issue with this algorithm is that we have to consume fresh copies of database states for each equality test, which is {\em unsustainable} for a database system that is designed to answer an unlimited number of queries.

\subsection{Search and Join}
In the classical data setting, given a set of objects $X = \{p_1, \ldots, p_n\}$ and a query object $q$, the search operation returns some $p_i \in X$ such that $p_i = q$ if such $p_i$ exists, and $\emptyset$ otherwise.  In the quantum setting, again due to the difficulty of distinguishing two quantum states within a distance of $\eps$, we propose the following approximation version.

\begin{definition}[$(\veps, \beta)$-search]
	Given a query state ${\phi}$ and a database $X$, if there exist a state ${\psi} \in X$ such that $D({\phi}, {\psi}) \le \veps$,  return a state ${\psi'} \in X$ with $D({\phi}, {\psi'}) \le \beta \veps$.  Otherwise, either return a state ${\psi'} \in X$ with $D({\phi}, {\psi'}) \le \beta \veps$ or return $\emptyset$.  
\end{definition}
In other words, if there exists a state in the database which has a trace distance no more than $\veps$ from the query state $\phi$, we return a state in $X$ whose distance is no more than $\beta\veps$ from $\phi$ (similar to the ANN search). Else if all states in the database have distances larger than $\beta\veps$ from the query state, we return $\emptyset$.  In other cases, we either return a database state with distance no more than $\beta\veps$ from the query state or return $\emptyset$. 

The most straightforward way is to perform the $(\veps, \beta)$-equality-test for each database state ${\psi} \in X$ with the query state ${\phi}$.  By the above algorithm for equality test (setting $\delta = 1/m^2$), we can determine with probability $(1 - m \delta) = (1 - o(1))$   whether there exists a state ${\psi} \in X$ such that the $(\veps, \beta)$-equality-test on ${\phi}$ and ${\psi}$ returns $1$.  The above procedure takes $O\left(m \log d\log m/\veps^2\right)$ quantum time, which is linear in terms of the number of states in the database.  Another significant limitation of this method is the necessity of using fresh copies of the database states for each search operation because of the equality test, making the database system unsustainable.
\smallskip

A closely related operation to search is join, which is one of the most important operations in relational database systems.  We introduce the quantum version of {\em natural join} as follows.

\begin{definition}[$(\veps, \beta)$-natural-join]
	Given two databases $X$ and $Y$ of quantum states, we want to output a set that includes all pairs of states $(\phi, \psi)\ (\phi \in X, \psi \in Y)$ such that $D(\phi, \psi) \le \eps$, and excludes all pairs $(\phi, \psi)$ such that $D(\phi, \psi) > \beta\eps$. The decisions for other pairs can be arbitrary.
\end{definition}

\subsection{Selection and Sorting}
\label{sec:selection}

In relational databases for classical data, selection is typically denoted by $\sigma_\theta(R)$, where $R$ is a relation and $\theta$ is a propositional formula that involves an attribute, a comparison operator in the set $\{<, >, \le, \ge, =, \neq\}$, and a constant value for comparison (e.g., age $\ge  8$). However, in the quantum data setting, quantum states cannot be directly compared. We can only apply a measurement $\M$ on the state ${\phi}$ and get a random outcome according to the distribution $\M({\phi})$. As a classical analog, we would say a person's age is $5$ with probability $0.6$ and $10$ with probability $0.4$.\footnote{This assembles probabilistic databases, but in the quantum data setting the probability distribution is not given explicitly, and the support size of the distribution is exponential in terms of the number of qubits of each quantum state.}  We thus look at the expectation value $\phi^\dagger M \phi$ for the observable $M$ corresponding to $\M$. 

The quantity $\phi^\dagger M \phi$ holds significant importance in quantum mechanics (see, e.g., the textbook~\cite{SC95}).  It can be used to provide an estimate of the system's average energy in a particular state, describe the level of non-classical correlations between entangled particles, quantify quantum information such as entropy, coherence, and entanglement, etc. 

We define the $\veps$-approximate `$\ge$' selection operation for quantum data as follows.
\begin{definition}[$(\eta, \veps)$-selection]
	\label{def:selection}
	Given a database $X$, an observable $M$, a threshold $\eta$, and an error parameter $\veps$, return a set of states $S \subseteq X$ such that $S$ includes all database states ${\phi}$ such that $\phi^\dagger M \phi \ge \eta$, but excludes all ${\phi}$ such that $\phi^\dagger M \phi \le \eta - \veps$.
\end{definition}

Note that the $\veps$-approximate equality selection can be implemented by taking the difference between $(\eta - \veps, \veps)$-selection and $(\eta + 2 \veps, \veps)$-selection, which includes all ${\phi}$ with $\eta - \veps \le \phi^\dagger M \phi \le \eta + \veps$ and excludes all ${\phi}$ with $\phi^\dagger M \phi \le \eta - 2\veps$ or $\phi^\dagger M \phi \ge \eta + 2\veps$.  In the  context of approximation, we can consider `$<$' and `$>$'  the  same as `$\le$' and `$\ge$', respectively.

We also note that $(\veps, \beta)$-search can also be handled by looking at $\phi^\dagger M \phi$ for a specific observable $M$, although this solution is not as efficient as that using the particular sketches that we shall design for the search operation. We have included a reduction from $(\eps, \beta)$-search to $(\eta, \eps)$-selection in Appendix~\ref{app:search-as-selection}.

In the context of databases, we are particularly interested in the following type of observables. 
\revone{\begin{definition}[$k$-local observable] 
\label{def:local-ob}
An observable $O$ of a system with $n$ qubits is called \(k\)-local if it can be written as a sum of a constant number of terms, each acting on at most \(k\) qubits. For instance, a $2$-local observable in a $3$-qubit system might look like:
\[
O = O_{12} \otimes I_3 + I_1 \otimes O_{23},
\]
Where \(O_{12}\) and \(O_{23}\) are operators acting on the pairs of qubits (1,2) and (2,3) respectively, while \(I_3\) and \(I_1\) are the identity operators acting on the remaining qubits.
\end{definition}
}

$k$-local observables have been well studied in the literature (see \cite{CNY23,KKR06} and references therein).  They are interesting because, in most practical scenarios, our goal is to identify specific properties   of a quantum state (e.g., the energy, momentum, or spin of a photon) that rely on a small subset of qubits of the state.   This is similar to the classical setting where most queries depend on a few attributes of a relational database table. For example, suppose we want to retrieve all records in a table containing patient information for individuals aged $80$ years or older with  systolic blood pressure at least $140$, we only need to look at two attributes in the table: age and blood pressure.  If we view each qubit of a quantum state as an attribute (e.g., spin, position, momentum, polarization, etc.), then a $k$-local observable performs selection on at most $k$ attributes of the quantum state. 

\smallskip

A related problem of selection is sorting. \rev{As a motivation, we would like to sort a set of given quantum states according to their average energy with respect to an observable determined by a particular application. Note that there is no natural order between the quantum states themselves. Therefore, introducing an observable and computing the expectation value is somewhat necessary to establish a total order between the quantum states.}

We define the sorting operation with respect to an observable $M$ as follows. Similar to the selection operation, we introduce an additive approximation $\veps$ in the sorted order.
\begin{definition}[$\veps$-sorting]
	\label{def:sorting}
	Given a database $X$ of $m$ states, an observable $M$, and an error parameter $\veps$, return an order $(\phi_1, \phi_2, \ldots, \phi_m)$ of the states in $X$ such that for all $i = 1, \ldots, m-1$, we have ${\phi_i}^\dagger M\phi_i \le {\phi_{i+1}}^\dagger M\phi_{i+1} + \veps$. 
\end{definition}

%% file: data.tex
\section{Sketches for Quantum Data Operations}
\label{sec:sketch}

In this section, we introduce two quantum data sketches, {\em vector sketches} and {\em shadow seeds}, which are summaries of the original states for efficiently handling previously mentioned database operations. 

Before delving into the details, let us use metaphors to provide some very high-level intuition of the two data summarizing methods. The vector sketches can be seen as capturing snapshots of the state from different angles, while each shadow seed can be seen as a piece of information gleaned from the state. Using multiple shadow seeds, we can reconstruct the original state at varying levels of resolution.

\subsection{Vector Sketches for Equality-Test, Search, and Join}
\label{sec:vector}

The concept of vector sketch is to represent a quantum state ${\phi}$ as a vector in $\mathbb{R}^t$ with $t \ll d$ instead of a vector in $\mathbb{C}^d$, while preserve certain distance properties.   
In this section, we design vector sketches for quantum states and then use them to conduct equality test, search, and join.

A natural way to construct the sketch is to take a number of random measurements on ${\phi}$, and write down the measurement outcomes as a vector.  
The following result is due to Sen~\cite{Sen06}, rewritten for pure quantum states. 
\begin{theorem}[\cite{Sen06}]
	\label{thm:Sen}
	Let ${\phi}$ and ${\psi}$ be two pure quantum states in $\mathbb{C}^d$.  With probability at least $\left(1 - e^{-\Omega(d)}\right)$ over the choice of a random  measurement basis $\M_d = \{M_1, \ldots, M_d\}$, there exists a universal constant $c \in (0, 1)$ such that
	\begin{equation}
	\label{eq:L1}
		c \cdot D({\phi}, {\psi}) \le \norm{\M_d({\phi}) - \M_d({\psi})}_1 \le  D({\phi}, {\psi}).
	\end{equation} 
\end{theorem} 

Theorem~\ref{thm:Sen} connects the trace distance of two quantum states to the $\ell_1$ distance of their measurement outcome distributions.  We note that the {\em distortion} in (\ref{eq:L1}), $D({\phi}, {\psi})/(c D({\phi}, {\psi})) = 1/c$,  is a big constant whose value left unspecified in \cite{Sen06}.

Vectors $\M_d({\phi})$ and $\M_d({\psi})$ are discrete distributions with outcomes $\{1, 2, \ldots, d\}$. It is well-known that for a discrete distribution $\mu$ over a domain of size $d$, using $\Theta\left({(d + \log(1/\delta))}/{\eps^2}\right)$ samples we can obtain an empirical distribution $\widetilde{\mu}$ such that $\norm{\mu - \widetilde{\mu}}_1 \le \eps $ with probability $1 - \delta$ (see, e.g., \cite{Canonne20}).  

\begin{corollary}
	\label{cor:Sen}
	Let $\widetilde{\M_d}({\phi})$ and $\widetilde{\M_d}({\psi})$ be the empirical distributions of measurement outcomes by applying $\M_d$ in Theorem~\ref{thm:Sen} to $c_s (d + \log(1/\delta))/\eps^2$ (for a sufficiently large constant $c_s$) copies of ${\phi}$ and ${\psi}$, respectively.  With probability $1 - \delta - e^{-\Omega(d)}$, we have 
	\begin{equation*}
		c \cdot D({\phi}, {\psi}) - \eps \le \norm{\widetilde{\M_d}({\phi}) - \widetilde{\M_d}({\psi})}_1 \le D({\phi}, {\psi}) +  \eps,
	\end{equation*} 
where $c \in (0, 1)$ is a universal constant.
\end{corollary}

We can view $\widetilde{\M_d}({\phi})$ and $\widetilde{\M_d}({\psi})$ as two empirical probability vectors.  However, since $d = 2^n$ for a $n$-qubit state, it is both space-expensive to store $\widetilde{\M_d}({\phi})$ and time-expensive to use it for database operations.

\paragraph{Embedding to $L_1$-space} We aim to address the issue of efficiency in both time and space by showing that there is another distribution of measurements whose number of outcomes is {\em independent} of the state dimension $d$, for which a similar connection exists between the trace distance of two quantum states and the $\ell_1$ distance of the corresponding measurement outcome distributions.  Moreover, the distortion of our sketching can be made arbitrarily close to $1$ (compared with $1/c$ in \eqref{eq:L1}).  It is worth noting that this distortion will significantly impact the efficiency of the search and join operations, as we will discuss shortly.

Our result is summarized in the following theorem. 


\begin{theorem}
	\label{thm:sketch-L1}
	Let ${\phi}$ and ${\psi}$ be two pure $d$-dimensional quantum states.  For any $\iota > 0$, there is a distribution $\pi$ of measurements with $k = c {\log(1/\delta)}/{\iota^2}$ outcomes for a sufficiently large constant $c$, such that a measurement $\M_k$ sampled randomly from $\pi$ satisfies
	\begin{equation*}
		(1 - \iota)D({\phi}, {\psi}) \le \sqrt{\frac{d}{k}} c_\tau  \norm{\M_k({\phi}) - \M_k({\psi})}_1  \le (1 + \iota)D({\phi}, {\psi})
	\end{equation*}
	 with probability at least $(1-\delta)$, where $c_\tau  \in [0.48, \sqrt{2}]$ is a universal computable constant.
    \rev{Additionally, the measurement sampling can be completed in $O(\log^8 d)$ time, and the sampled measurement can be represented as a quantum circuit with a gate complexity of $O(\log^2 d)$.}
\end{theorem} 

\rev{
\begin{proof}[Proof Overview]
At a high level, our approach leverages form dimension reduction through quantum measurements. We make use of a technique called  \textit{pretty good measurement} \cite{Holevo78} to generate random projective quantum measurements $\M$ with $k$ outcomes. The output of these measurements are random vectors serving as the embedding of the state $\phi$ into $\mathbb{R}^k$. 

We start by picking a random basis for $\mathbb{C}^d$ based on the Haar measure \cite{Holevo13}. Let $x_t, y_t\ (t = 1, \ldots, d)$ be independent Gaussian random variables with mean zero and variance $\sigma^2=\frac{1}{2d}$, and let  $g \triangleq (c_1, \ldots, c_d) \in \mathbb{C}^d$ be a random vector where $c_t=x_t + i y_t$. We repeat this process and generate $d$ {\em complex} Gaussian random vectors $g_1, \ldots, g_d$.
These vectors are linearly independent with probability one; but they are {\em not} necessarily orthonormal.  We make use of pretty good measurement to orthogonalize and normalize these vectors. More precisely, we construct the operator (matrix)  $\Gamma \triangleq \sum_{t \in [d]} g_t^\dagger g_t$, and define the vector $\gamma_t \triangleq \Gamma^{-1/2} g_t$ for each $t \in [d]$.  We can show that $\gamma_1, \ldots, \gamma_d$ are linearly independent and are orthonormal. Moreover, the distribution of $\gamma_t$ is unitary invariant, and hence the Haar measure. Intuitively, $\gamma_t$ is distributed uniformly over surface of the unit sphere in $\CC^d$. Next, we randomly group ${\gamma_t}$'s into $k$ groups and form random projection operators as 
\begin{equation*}
    \Pi_j = \sum_{\ell \in [d/k]} (\gamma^{j}_\ell)^\dagger\gamma^{j}_\ell\ . \quad\quad (j = 1, \ldots, k)
\end{equation*}
Let $\M_k = \{\Pi_1, \cdots, \Pi_k\}$ be the corresponding measurement. Clearly, $\M$ is a valid measurement with probability one. This random measurement facilitates an embedding  of the quantum states in $\CC^d$ into $\RR^k$. We carefully analyze the distortion of the embedding (i.e., the  outcome distribution by applying $\M_k$ to the quantum state)
using tools from the concentration of measures and properties of the Haar distribution.  We show that the distortion of this embedding is no more than $(1+\iota)$ with probability $(1-\delta)$ when $k = c\log(1/\delta)/\iota^2$ for a constant $c$. The complete proof can be found in Appendix~\ref{sec:proof-thm-sketch-L1}.

The measurement construction described above could require polynomial time in $d$. However, we demonstrate that it can be sampled more efficiently from the Clifford group in classical time $O(\log^8 d)$, leveraging the properties of unitary 2-designs from quantum information theory. The details are provided in Appendix~\ref{sec:efficient-impl}.
\end{proof}
}

To approximate $\sqrt{\frac{d}{k}} c_\tau  \norm{\M_k({\phi}) - \M_k({\psi})}_1$ up to an additive error $\eps$, we have to approximate $\norm{\M_k({\phi}) - \M_k({\psi})}_1$ up to $\eps' = \frac{\eps}{\sqrt{d/k} \cdot c_\tau}$.  
We have the following immediate corollary.  

\begin{corollary}
\label{cor:sketch-L1}
For any $\iota > 0$, let $k = c \log(1/\delta)/{\iota^2}$ for a sufficiently large constant $c$, and let $\widetilde{\M_k}({\phi})$ and $\widetilde{\M_k}({\psi})$ be the empirical distributions of the outcomes by applying independent random measurements $\M_k$ in Theorem~\ref{thm:sketch-L1} to $c_s d/\eps^2$ (for a sufficiently large constant $c_s$) copies of ${\phi}$ and ${\psi}$, respectively.  With probability at least $1 - \delta$, we have
\begin{align*}
	(1-\iota) D({\phi}, {\psi}) - \eps \le \sqrt{\frac{d}{k}} c_\tau \norm{\widetilde{\M_k}({\phi}) - \widetilde{\M_k}({\psi})}_1  \le (1+\iota)D({\phi}, {\psi}) +  \eps,
\end{align*} 
where $c_\tau \in [0.48, \sqrt{2}]$ is the same constant in Theorem~\ref{thm:sketch-L1}.
\end{corollary}

\paragraph{Embedding to $L_2$-space}  The sketch we have constructed for the $L_1$-space can also be applied to the $L_2$-space, albeit through a different analysis. The $\ell_2$ distance is interesting since we know from Theorem~\ref{thm:LSH} that $\ell_2$ enjoys a slightly better ANN scheme in term of time and space complexities, which will be useful for speeding up search and join operations.  The proof of the following theorem can be found in Appendix~\ref{sec:proof-thm-sketch-L2}.
\begin{theorem}
	\label{thm:sketch-L2}
	Let ${\phi}$ and ${\psi}$ be two pure $d$-dimensional quantum states.  For any $\iota > 0$, there is a distribution $\pi$  of measurements with $k = c \log(1/\delta)/\iota^2$ outcomes for a sufficiently large constant $c$, such that a measurement $\M_k$ sampled randomly from $\pi$ satisfies
	\begin{equation*}
		 (1 - \iota)D({\phi}, {\psi}) \le \sqrt{\frac{d}{2}} \norm{\M_k({\phi}) - \M_k({\psi})}_2 \le  (1 + \iota)D({\phi}, {\psi})
	\end{equation*} 
with probability at least $1 - \delta$. \rev{Additionally, the measurement sampling can be completed in $O(\log^8 d)$ time, and the sampled measurement can be represented as a quantum circuit with a gate complexity of $O(\log^2 d)$.}
\end{theorem}

For a discrete distribution $\mu$ over a domain of size $d$ for any $d \ge 1$, it takes $\Theta\left({\log(1/\delta)}/{\eps^2}\right)$ samples to obtain an empirical distribution $\widetilde{\mu}$ such that $\norm{\mu - \widetilde{\mu}}_2 \le \eps$ with probability $1 - \delta$ (see, e.g., \cite{Canonne20}).  We have the following corollary.

\begin{corollary}
	\label{cor:sketch-L2}
	For any $\iota > 0$, let $k = c {\log(1/\delta)}/{\iota^2}$ for a sufficiently large constant $c$, and let $\widetilde{\M_k}({\phi})$ and $\widetilde{\M_k}({\psi})$ be the empirical distributions of the outcomes by applying independent random measurements $\M_k$ in Theorem~\ref{thm:sketch-L2} to $c_s {d \log(1/\delta)}/{\eps^2}$ (for a sufficiently large constant $c_s$) copies of ${\phi}$ and ${\psi}$, respectively.  With probability $1 - \delta$, we have
	\begin{align*}
		(1 - \iota)D({\phi}, {\psi}) - \eps \le \sqrt{\frac{d}{2}} \norm{\widetilde{\M_k}({\phi}) - \widetilde{\M_k}({\psi})}_2 \le (1 + \iota)D({\phi}, {\psi}) + \eps.
	\end{align*} 
\end{corollary}

\rev{\noindent{\bf Johnson-Lindenstrauss Lemma in Our Context.} \
It is natural to ask whether existing dimension reduction techniques, such as the Johnson–Lindenstrauss (JL) lemma, can be applied directly to the $d$-dimensional vector representation $\alpha(\phi) = (\alpha_1, \ldots, \alpha_d) \in \mathbb{C}^d$ of a quantum state $\phi$, or the outcome distribution $p(\phi) = (p_1, \ldots, p_d) \in \mathbb{R}^d \ (p_i = \abs{\alpha_i}^2)$ when measured in the computational basis. After all, we can use quantum tomography to learn the representation $(\alpha_1, \ldots, \alpha_d)$ approximately. We would like to first point out that a direct application will not work, since we can construct simple examples demonstrating inherent distortions between the trace distance of quantum states and the \(\ell_1/\ell_2\) distances of their \(d\)-dimensional vector representations (\(\alpha(\phi)\) or \(p(\phi)\)), even when all the coordinates are real-valued and before any dimension reduction step. We leave the detailed examples and calculation to Appendix~\ref{app:distortion}. In our examples, for the $\alpha(\phi)$ vector representation, the distortions between the trace distance of quantum states and the $\ell_1$ and $\ell_2$ distances of the two corresponding vectors are at least $\sqrt{{d}/{6}}$ and $\sqrt{1.5}$, respectively. And for the $p(\phi)$ vector representation, the distortions between the trace distance of quantum states and the $\ell_1$ and $\ell_2$ distances of the two corresponding vectors are at least $\sqrt{3}$ and $\sqrt{{3d}/{4}}$, respectively. Moreover, the JL lemma only takes real vectors. 

We also note that there exists a near-linear lower bound for dimension reduction in the $L_1$ space~\cite{ACNN11}, indicating that, unlike the JL lemma for $L_2$ space, dimension reduction in the $L_1$ space is not generally possible.

We note that there is a way to circumvent the issues for embedding quantum states into the $L_2$ space: for each state $\phi$, we write its density matrix $\phi\phi^{\dagger}$ as a real-valued $2d^2$ dimensional vector $v_\phi$. By some calculation, we can show that the $\ell_2$ distance of $v_\phi$ and $v_\psi$ preserves the trace
distance of the two original pure states $\phi$ and $\psi$. We then perform dimension reduction on the vectors $v_\phi$ using the JL lemma. Our sketching algorithm has the following advantages compared with this ``full tomography plus JL lemma'' approach (setting the error probability $\delta = 0.01$):
\begin{enumerate}
    \item The memory usage of our sketch construction is {\em independent} of $d$, while the memory needed for storing the classical vector representation of the quantum state $\phi$ is $O(d)$ and that for the density matrix $\phi\phi^{\dagger}$ is $O(d^2)$. 

    \item Our sketch construction takes  $\tilde{O}(d /\eps^2)$ time, while the full (pure) quantum state tomography takes  $O(d^2/\eps^5)$~\cite{FBK21} time and the dimension reduction using the JL lemma needs another $O(d^2/\eps^2)$ time.
\end{enumerate}
These comparisons demonstrate that our sketch construction using direct quantum measurements significantly outperforms the method of first converting the quantum state to its classical description followed by dimension reduction, both in terms of time and space, which are the main focus of this paper.
}

\smallskip

We now apply our embedding results to database operations.

\paragraph{The Equality-Test Operation}
We observe that Corollary~\ref{cor:sketch-L1} and Corollary~\ref{cor:sketch-L2} directly provide a way for solving $(\eps,\beta)$-equality-test. We just set $\iota = \eps = \frac{\veps}{2}$, and use the $\ell_1$ or $\ell_2$ distances between the two vector sketches $\widetilde{\M_k}({\phi})$ and $\widetilde{\M_k}({\psi})$ to estimate $D(\phi, \psi)$ up to an additive error $\veps$ with probability $1-\delta$. The running time is bounded by $O(k) = O(\log(1/\delta)/\veps^2)$. 

\paragraph{The Search Operation}
We now illustrate how to use vector sketches and approximate nearest neighbor (ANN) to perform $(\eps, \beta)$-search on quantum states.

Let $\eps$ and $(1+\iota)$ be the additive error and multiplicative error in Corollary~\ref{cor:sketch-L1}/Corollary~\ref{cor:sketch-L2} for building $\left\{\left.\widetilde{\M_k}({\phi})\ \right|\ {\phi} \in X\right\}$, respectively. We assume that an LSH indexing structure has already been built on top of $\widetilde{\M_k}({\phi})$'s to achieve the time and space usages stated in Theorem~\ref{thm:LSH}.   To handle $(\veps, \beta)$-search, we call 
$\ANN\left(\widetilde{\M_k}({\phi}), \left\{\left.\widetilde{\M_k}({\psi})\ \right|\ {\psi} \in X \right\}, (1+\iota)\veps, \beta_{nn} \right)$,
where $\beta_{nn} = {\beta}/{(1+\iota + \eps/\veps)}$ is the parameter for the tradeoff between the distortion and the time/space complexity in ANN.
By Corollary~\ref{cor:sketch-L1}/Corollary~\ref{cor:sketch-L2} and Theorem~\ref{thm:LSH}, if there exists a state ${\psi} \in X$ such that $D({\phi}, {\psi}) \le \veps$, then ANN returns a state ${\psi'} \in D$ such that $D({\phi}, {\psi'}) \le \beta \veps$.  On the other hand, ANN either returns a state ${\psi'} \in D$ with $D({\phi}, {\psi'}) \le \beta \veps$, or returns $\emptyset$. 

By Theorem~\ref{thm:LSH}, it takes $O(k m^\gamma) = O(m^\gamma \log m / \veps^2)$ classical time to perform the search. The space for storing the LHS index is $O(km + m^{1+\gamma}) = O(m\log m/\veps^2 + m^{1+\gamma})$, where $\gamma \approx 1/\beta_{nn}$ for $\ell_1$ and $\gamma \approx 1/\beta_{nn}^2$ for $\ell_2$.  

We note that in the above approach, we have to make sure that $\beta_{nn} \ge 1$.  In other words, we can only handle $(\eps, \beta)$-search with $\beta \ge (1+\iota +  \eps/\veps)$. However, since $\eps$ and $\iota$ can be positive constants arbitrarily close to $0$, we can essentially handle all constants $\beta > 1$. Certainly, the higher the value of $\beta$, the larger $\beta_{nn}$ that we can pick for reducing the query time and space usage in the ANN search. In practice, a reasonably large constant $\beta$ may be okay, as the trace distance between two quantum states that are generated by separate entities or experiments is typically much larger than that between two states originating from the same entity or experiment (due to quantum noise or preparation errors).  

Setting $\delta = 1/m^2$, $\iota = 0.01$ and $\eps = 0.01\veps$, we have $\beta_{nn} \ge 0.98\beta$, and consequently $\gamma \le 1.05/\beta^2$. Applying our vector sketch with respect to the $\ell_2$ distance and the corresponding ANN search, we have the following theorem.
\begin{theorem}
	\label{thm:search}
	There is an index of size $O\left(\frac{m\log m}{\veps^2} + m^{1+\frac{1.05}{\beta^2}}\right)$, using which we can solve $(\eps, \beta)$-search on a database of $m$ quantum states with success probability $1-o(1)$ and classical time $O\left(m^{\frac{1.05}{\beta^2}} \cdot \frac{\log m}{\veps^2}\right)$.
\end{theorem}

Note that the index space cost is {\em independent} of $d$, and the query time is {\em sublinear} in $m$ (for $\beta > \sqrt{1.05}$) and {\em independent} of the state dimension $d$.

\paragraph{The Join Operation}  The sketch-based approach can also be used for join. Given a set of sketch vectors $\left\{\left.\widetilde{\M_k}({\phi})\ \right|\ {\phi} \in X \right\}$, we can apply the same hashing process as that for the ANN search, and then verify (by computing the actual distance) all pairs of vectors that collide on at least one hash function.  The space cost is the same as that of the search. The query time is dependent on the size of the join output, but it is still independent of the state dimension
$d$.

\subsection{Shadow Seeds for Selection and Sorting}
\label{sec:shadow}
\input{CST}

\paragraph{The Selection Operation}
It is easy to see that Theorem~\ref{thm:QSS} directly implies an algorithm for handling $(\eta, \veps)$-selection: Setting $\delta = 1/m^2$, we can estimate $\phi^\dagger M \phi$ up to an additive error $\veps$ with probability $(1 - 1/m^2)$ for each $n$-qubit database state ${\phi}$ using an $N \times n$ shadow seed matrix, where $N \ge 9^{k} \norm{M}_\infty^2 \cdot {2\log m}/{\veps^2}$.  By a union bound over $m$ database states, we can solve the $(\eta, \veps)$-selection problem with probability $(1 - 1/m)$.  The query time is bounded by $N m \cdot \text{poly}(k) = 9^k  m \log m \cdot \text{poly}(k) \norm{M}_\infty^2 /{\veps^2}$.

\begin{theorem}
	\label{thm:selection}
	There is an index of size $O\left(9^{k} n W^2 {\log m}/{\veps^2}\right)$, using which we can solve for any $k$-local observable $M\ (\norm{M}_\infty \le W)$ the $(\eta, \eps)$-selection on a database of $m$ $n$-qubit quantum states with success probability $(1-o(1))$ and quantum time $9^k m{\log m} W^2 \text{poly}(k) /{\veps^2}$.
\end{theorem}

\paragraph{The Sorting Operation}  Since the shadow seed matrix can be used for estimating the expectation value $\phi^\dagger M \phi$ up to an additive error $\veps$, we can use it for $\veps$-sorting with the same space and time complexity as that for the selection operation.

%% file: CST.tex
In this section, we develop a classical data summarization that  can be used to estimate the expectation value $\phi^\dagger M \phi$ for an arbitrary $k$-local observable $M$. 
\rev{We make use of the classical shadow tomography (CST), introduced in \cite{HKP20}, to approximate $\phi^\dagger M \phi$ up to a small additive error. CST tries to extract minimal information about the quantum state, without performing complete tomography, to estimate certain properties of the state described by observables. 

For completeness, let us briefly describe the CST procedure using Pauli measurements. For each of the $N$ copies of ${\phi}$, we select $n$ unitary operators, $U_1, \ldots, U_n$, randomly and independently from the set $\{I, H, S^\dagger H\}$, where $H$ is the Hadamard gate and $S=\sqrt{Z}$ is the square root of the Pauli-$Z$ gate; see Appendix~\ref{app:gate} for their matrix representations. We then apply $U_j$ to the $j$-th qubit of ${\phi}$ and measure the state on the computational basis. The result is a binary string  $b_1, \ldots, b_n \in \{0, 1\}$. 
The $n$ pairs $\{b_j, \mathtt{index}(U_j))\}_{j = 1}^n$ form a row vector, where $\mathtt{index}(U_j)$ is the index of $U_j$ in the set $\{I, H, S^\dagger H\}$. We then repeat this process for $N$ times, getting $N$ rows, forming the seed matrix $A({\phi}) = \{b_{i,j}, \mathtt{index}(U_{i,j})\}_{i \in [N], j \in [n]}$.
We call $A({\phi})$ the {\em shadow seeds}.
Clearly, $A({\phi})$ can be stored using $O(n N)$  classical bits, since each entry of $A({\phi})$ belongs to $\{0,1\}\times \{0,1,2\}$.  
}

At the time of query, given a $k$-local observable $M$, we first construct $k$-local classical shadows  ${\tilde{\phi}_i}$ of the database state ${\phi}$ from each row $i \in [N]$ of its seed matrix $A({\phi})$ with respect to the $k$-local observable $M$. Suppose $M$ depends non-trivially on the $k$ qubits indexed by $Q \triangleq \{q_1, \ldots, q_k\}$. \rev{Let $e_0 = (0, 1)^T, e_1=(1,0)^T$ be the standard basis vectors in the two dimensional plane. For each row $i\in [N]$ and column $j\in Q$, we first construct a vector ${v_{i, j}} = U_{i, j} e_{b_{i, j}}$.  Next, we construct the $i$-th shadow as a $2^k\times 2^k$ matrix 
$\hat{\rho}_i = \bigotimes_{j\in Q} \qty(3 v_{i,j} v^\dagger_{i,j} - I),$
where $I$ is the $2 \times 2$ identity matrix.
Finally, the estimator for $\phi^\dagger M \phi$ is given by 
$T = \frac{1}{N} \sum_{i \in [N]} \tr{ M \hat{\rho}_i}$.
The following theorem states that $T$ is a good approximation of the expectation value $\phi^\dagger M \phi$.

\begin{theorem}[Based on \cite{HKP20}]
	\label{thm:CST}
	The above procedure prepares an $N \times n$ shadow seed matrix $A({\phi})$ given $N$ copies of an $n$-qubit quantum state ${\phi}$, such that for any given $k$-local observable $M$, if $N \ge 4^k \norm{M}_\infty^2 {\log(1/\delta)}/{\veps^2}$,
	the estimator $T$ approximates $\phi^\dagger M \phi$ up to an additive error 
	$\veps$ with probability $(1 - \delta)$ using $A({\phi})$.  Moreover, the time for computing $\phi^\dagger M \phi$ using $A({\phi})$ is bounded by $O\left(2^{2k} N\right) (\propto 16^k)$, and the space for storing $A({\phi})$ is $O(N n)$ classical bits.
\end{theorem}

Note that the space cost and query time are both independent of the state dimension $d$.
\smallskip

Typically, the $k$-local observable $M$ can be expressed as a quantum circuit with $\text{poly}(k)$ gate complexity. In this case,
we propose a new estimation algorithm to further improve the total query time from $O(16^k)$ to $O(9^k)$ (omitting other less critical factors) by an approach we call QCQC (quantum$\to$classical$\to$quantum$\to$classical). We have the following theorem, whose proof can be found in Appendix~\ref{app:QSS}.
}

\rev{
\begin{theorem}
	\label{thm:QSS}
	There is a procedure for preparing an $N \times n$ shadow seed matrix $A({\phi})$ given $N$ copies of an $n$-qubit quantum state ${\phi}$, such that for any given $k$-local observable $M$ with $\text{poly}(k)$ gate complexity, if $N \ge 9^k \norm{M}_\infty^2 {\log(1/\delta)}/{\veps^2}$,
	we can approximate $\phi^\dagger M \phi$ up to an additive error 
	$\veps$ with probability $(1 - \delta)$ using $A({\phi})$.  Moreover, the quantum time for computing $\phi^\dagger M \phi$ using $A({\phi})$ is bounded by $O\left(N \text{poly}(k) \right) (\propto 9^k)$, and the space for storing $A({\phi})$ is $O(N n)$ classical bits.
\end{theorem}
}

%% file: future.tex
\section{Conclusion and Future Work}
\label{sec:future}

In this paper, we have defined basic database queries for quantum data and proposed several classical sketches of quantum states to facilitate these queries. We consider our work a preliminary step towards a comprehensive quantum data management system. Numerous questions and directions remain open following this work. 
We list a few below.

\paragraph{Support More Data Operations}
This paper primarily focuses on two basic database operations: search and selection, along with several related operations. We would like to expand the support to more complex operations for data analytics, such as  {\em clustering} and {\em classification}, for which we may need to develop new classical summaries of the quantum states for the sake of efficiency. 

\paragraph{Mixed States}
In various scenarios, such as when the description of a quantum system is unknown due to quantum noise, the use of a density operator (or, density matrix) for describing {\em mixed} quantum states becomes more convenient. Suppose the quantum system is in one of a collection of $d$-dimensional pure states $\{{\phi_1}, \ldots, {\phi_k}\}$, we can represent a mixed quantum state as 
$\rho = \sum_{i=1}^k p_i \phi_i {\phi_i}^\dagger$,
where $p_1, \ldots, p_k \ge 0$ and $\sum_{i=i}^k p_i = 1$.  We can view $\rho$ as a convex combination of outer products of pure states ${\phi_i}$, where each  $\phi_i{\phi_i}^\dagger$ is associated with a probability $p_i$. We anticipate that results presented in this paper can be extended to mixed states, although the technical aspects of this generalization require further investigation.

\paragraph{The Integration with the Theory of Relational Databases}  A key feature of our proposed model is that quantum data is represented entirely in the classical format. This unique aspect enables us to integrate our model with established theories related to indexing, query execution, and query optimization in relational databases designed for classical data. However, the integration process will likely require the redesign of multiple components to accommodate the inherent differences stemming from the distinct definitions of database operations for quantum data.

%% file: appendix.tex
\section{More Preliminaries}
\label{sec:pre-full}

\subsection{Basics of Quantum Information (A More Formal Approach)}
\label{sec:quantum-basic}
\vspace{2mm}
\noindent{\bf Quantum States and Qubits.}
We can represent a $d$-dimensional pure quantum state as a vector in $\mathbb{C}^d$.  Using the standard Dirac bra-ket notation, we write
$
\ket{\phi} = \sum_{i=0}^{d-1} \alpha_i \ket{i},
$
where $\ket{0}, \ket{1}, \ldots, \ket{d-1}$ is an orthonormal basis in  $\mathbb{C}^d$ (referred as the {\em computational basis}), and $\alpha_i$'s are called {\em amplitudes} with the property $\sum_{i=0}^{d-1} \abs{\alpha_i}^2 = 1$.

Let $\bra{\phi}$ denote the conjugate transpose of $\ket{\phi}$, and let  $\braket{\phi}{\varphi}$ and $\ketbra{\phi}{\varphi}$ denote the inner product  and outer product of vectors $\ket{\phi}$ and $\bra{\varphi}$, respectively.  

A {\em qubit} is a $2$-dimensional quantum state and can be represented as $\ket{\phi} = \alpha_0 \ket{0} + \alpha_1 \ket{1}$, where $\abs{\alpha_0}^2 + \abs{\alpha_1}^2 = 1$.  Generally, a $n$-qubit state can be represented as 
$$\ket{\phi} = \sum_{x_1 \cdots x_n \in \{0,1\}^n} \alpha_{x_1 \cdots x_n} \ket{x_1 \cdots x_n},$$
where $\{\ket{x_1 \cdots x_n}\}$ are the computational basis states of the $n$-qubit system, and it holds that $\sum_{x_1 \cdots x_n \in \{0,1\}^n} \abs{\alpha_{x_1 \cdots x_n}}^2 = 1$.

A quantum state is called {\em separable} if it can be written as a tensor product of at least two states
$\ket{\phi} = \ket{\phi_1} \otimes \ket{\phi_2} \otimes \cdots \otimes \ket{\phi_k}$,
which is often abbreviated as $\ket{\phi_1} \ket{\phi_2} \cdots \ket{\phi_k}$, or $\ket{\phi_1 \phi_2 \cdots \phi_k}$.
Otherwise, the state is called {\em entangled}.  A classical entangled state is the {\em Bell state} $\ket{\phi} = \frac{\ket{00}+\ket{11}}{\sqrt{2}}$.

\paragraph{Quantum Operations}  There are two types of quantum operations.  The first is called {\em unitary transformation}. That is, we apply a unitary operator $U$ to a quantum state $\ket{\phi}$ and get $U\ket{\phi}$.\footnote{A unitary operator is a linear operator $U$ such that $U U^\dagger = U^\dagger U = I$.}  This type of operation is used to describe the evolution of a {\em closed} quantum system.   
The second type of operations are {\em measurements}. Quantum measurements are the interface to obtain classical information about quantum states.  Under the POVM  (Positive Operator Valued Measures) formalism, a quantum measurement $\M$ is described as a collection of $d \times d$ positive semi-definite operators $\{M_i\}$ with $\sum_i M_i = I$; each $M_i$ is associated with a measurement outcome $o_i$, which can be chosen by the experimentalist.  When performing $\M$ on a quantum state $\ket{\phi}$, the probability of getting the outcome $o_i$ is given by $\ev{M_i}{\phi}$.   
Let $\M(\ket{\phi})$ be the probability distribution of the measurement outcomes after applying $\M$ on $\ket{\phi}$.

A {\em projective measurement} is a special case of a POVM where $M_i$'s are projective operators, i.e., $M_i^2=M_i$. An example of a projective measurement is the measurement on the computational basis where $M_i=\ketbra{i}$.  Any POVM can be written as a unitary operator $U$ followed by a projective measurement.  

{\em Observables} are physical variables that can be measured. An observable is represented by a Hermitian operator $M$ whose eigenvalues are the set of possible outcomes. The observable spectrally decomposes as $M = \sum_i  \lambda_i M_i$, where $M_i$ represents the projector onto the eigenspace of $M$ associated with the eigenvalue $\{\lambda_i\}$. 
The observable $M$ can be associated  with the measurement $\M = \{M_i\}$ with outcomes $\{\lambda_i\}$. 
The expected value of an observable on a state $\ket{\phi}$ is expressed as $\E[\M(\ket{\phi})] = \ev{M}{\phi}$.

When we say a measurement is performed on a $d$-dimensional quantum state $\ket{\phi}$ in the computational basis, we mean that the measurement $\M = \{M_0, M_1, \ldots, M_{d-1}\}$ with $M_i = \ketbra{i}$ is applied on $\ket{\phi}$. It is noteworthy that the probability of observing the measurement outcome $o_i$,  denoted by $\ev{M_i}{\phi}$, is equal to $\abs{\braket{\phi}{i}}^2 = \abs{\alpha_i}^2$, wherein $\alpha_i$ is the $i$-th amplitude of the quantum state $\ket{\phi}$.

A crucial property of quantum mechanics is that each measurement would cause a disturbance to the quantum states.  If the measurement outcome is $o_i$, then the post-measurement state of $\ket{\phi}$ can be written as
$$
\ket{\phi'} = \frac{B_i \ket{\phi}}{\sqrt{\ev{B_i^\dag B_i}{\phi}}}\ ,
$$ 
where $B_i$ satisfies $B_i^\dag B_i =M_i$.
This particular phenomenon significantly complicates the quantum data management, as the quantum state cannot be entirely ``recycled'' following a measurement.

\paragraph{Trace Distance}
Given two quantum states $\ket{\phi}$ and $\ket{\psi}$, we define their trace distance to be 
\begin{equation*}
	D(\ket{\phi}, \ket{\psi}) = \frac{1}{2}\norm{\dyad{\phi} - \dyad{\psi}}_1 = \sqrt{1 - \abs{\braket{\psi}{\phi}}^2}.
\end{equation*}

The trace distance is the most widely used distance measure for quantum states in the literature. It also has a nice physical meaning:
Let $\M = \{M_i\}$ be a POVM, and let $p_i \triangleq \ev{M_i}{\phi}$ and $q_i \triangleq \ev{M_i}{\psi}$. That is, $p_i$ and $q_i$ are the probabilities of obtaining measurement outcome $o_i$ on $\ket{\phi}$ and $\ket{\psi}$, respectively.  We have
$
D(\ket{\phi}, \ket{\psi}) = \max_{\M} \sum_i \abs{p_i - q_i},
$
which implies that if two quantum states are close in terms of trace distance, then any measurement conducted on these quantum states will yield probability distributions that are close in terms of the total variance distance.  In other words, two quantum states that are close in terms of the trace distance are statistically indistinguishable under measurements.

\subsection{Some Basic Quantum States and Gates}
\label{app:gate}

We list the vector representations of some basic quantum states that we need to use in this paper.
\begin{itemize}
\item \(\ket{0}\):
\[
\ket{0} = \begin{pmatrix} 1 \\ 0 \end{pmatrix}
\]

\item \(\ket{1}\):
\[
\ket{1} = \begin{pmatrix} 0 \\ 1 \end{pmatrix}
\]

\item \(\ket{+}\):
\[
\ket{+} = \frac{1}{\sqrt{2}} \begin{pmatrix} 1 \\ 1 \end{pmatrix}
\]

\item \(\ket{-}\):
\[
\ket{-} = \frac{1}{\sqrt{2}} \begin{pmatrix} 1 \\ -1 \end{pmatrix}
\]

\item \(\ket{+i}\):
\[
\ket{+i} = \frac{1}{\sqrt{2}} \begin{pmatrix} 1 \\ i \end{pmatrix}
\]

\item \(\ket{-i}\):
\[
\ket{-i} = \frac{1}{\sqrt{2}} \begin{pmatrix} 1 \\ -i \end{pmatrix}
\]
\end{itemize}

We make use of two basic quantum gates:
\begin{itemize}
	\item Hadamard gate \[
	H = \frac{1}{\sqrt{2}} \begin{bmatrix}
		1 & 1 \\
		1 & -1
	\end{bmatrix} \ .
	\]
	It turns $\ket{0}$ to $(\ket{0}+\ket{1})/\sqrt{2}$, and turns $\ket{1}$ to $(\ket{0}-\ket{1})/\sqrt{2}$. 
	
	\item Phase gate  \[
	S = \begin{bmatrix}
		1 & 0 \\
		0 & i
	\end{bmatrix} \ .
	\]
	It leaves  $\ket{0}$  unchanged, and turns $\ket{1}$ to $i\ket{1}$.

\item Another ser of special unitary operators are the {Pauli operators} defined as:
\[
I = \begin{bmatrix}
1 & 0 \\
0 & 1
\end{bmatrix}, \quad
X = \begin{bmatrix}
0 & 1 \\
1 & 0
\end{bmatrix}, \quad
Y = \begin{bmatrix}
0 & -i \\
i & 0
\end{bmatrix}, \quad
Z = \begin{bmatrix}
1 & 0 \\
0 & -1
\end{bmatrix}.
\]

Here, \( I \) is the identity operator, \( X \) represents a bit-flip operation, \( Z \) represents a phase-flip operation, and \( Y \) combines both bit-flip and phase-flip with an imaginary phase factor.
\end{itemize}
\subsection{Clifford Group}\label{sec:clifford}
In this paper, we make use of a special class of quantum unitary operations called the Clifford group, which  is a fundamental construct in quantum information theory. The Clifford group consists of unitary operations that map Pauli operators to other Pauli operators. Furthermore, the Clifford group is known for its efficient classical simulation, as described by the Gottesman-Knill theorem \cite{MichaelA.Nielsen2010}. Any unitary in 
in this group can be implemented (up to a global phase factor) using a circuit with only Hadamard, Phase, and CNOT gates. The formal definition of the Clifford group is given below. 
\begin{definition}
The {Clifford group} \( \mathcal{C}_n \) on \( n \)-qubits is defined as the normalizer of the Pauli group \( \mathcal{P}_n \) under the action of conjugation, that is 
\[
\mathcal{C}_n = \{ U \in U(2^n) \, | \, U P U^\dagger \in \mathcal{P}_n, \, \forall P \in \mathcal{P}_n \},
\]
where \( \mathcal{P}_n = \langle iI, X_j, Y_j, Z_j \, | \, j = 1, \dots, n \rangle \) is the \( n \)-qubit Pauli group generated by the identity \( I \) and the Pauli matrices \( X, Y, Z \) on each qubit, along with the phase factor \( i \). 
\end{definition}
\subsection{Mathematical Tools}
\label{sec:tool}

\begin{lemma}[Hoeffding’s inequality]\label{lem:hoeffding}
	Let \(X_1, \dotsc, X_n \in [0, 1]\) be i.i.d.\ random variables and 
	$X = \frac{1}{n} \sum_{i = 1}^n X_i$.	Then 
	\begin{equation*}
		\Pr[\abs{X - \EE[X]} > t] \le 2 \exp(-2t^2n)\,.
	\end{equation*}
\end{lemma}

\begin{lemma}[Generic Chernoff bound]\label{lem:chernoff}
	For any $a\in \RR$ and  random variable $X$ with moment generating function $M_X(t):=\EE[e^{tX}]$,
	\begin{align*}
		\PP[X\geq a]\leq \inf_{t>0}M_X(t)e^{-ta}.
	\end{align*}
\end{lemma}

\begin{lemma}[Berry-Esseen theorem]
	\label{lem:berry-esseen}
	Let $X_1, X_2, ..., X_n$ be i.i.d.\ random variables with $\EE[X_i]=0, \EE\left[X_i^2\right]=\sigma^2<\infty$ and $\EE\left[X_i^3\right]=\rho<\infty$. If $Y_n :=\frac{1}{\sigma\sqrt{n}}\sum_j X_j$, then 
	\begin{equation*}
		\abs{\PP[Y_n \leq x] - \phi_N(x)}\leq \frac{c\rho}{\sigma^3 \sqrt{n}},
	\end{equation*}
	where $\phi_N(x)$ is the CDF of $N(0,1)$ and $c\leq \sigma^2$ is a constant. 
\end{lemma}

\section{Missing Proofs}
\label{app:proof}

\input{norm}

\input{QSS}

\section{Distortions Between The Trace Distance and $\ell_1/\ell_2$ Distances of Quantum States}
\label{app:distortion}

\begin{table}[t!]
    \centering
    \begin{tabular}{|c|c|c|c|}
        \hline
        & $\phi$ \& $\psi$ & $\ket{0}$ \& $\ket{1}$ & distortion w.r.t. $D$ \\
        \hline
        $D$ & $\sqrt{\frac{3}{4}}$ & $1$ & --  \\
        \hline
        $L_1$ & $\sqrt{\frac{d}{2}}$ & $2$ & $\sqrt{\frac{d}{6}}$ \\
        \hline
        $L_2$ & $1$ & $\sqrt{2}$ & $\sqrt{\frac{3}{2}}$ \\
        \hline
        $L'_1$ & $1$ & $2$ & $\sqrt{3}$ \\
        \hline
        $L'_2$ & $\sqrt{\frac{2}{d}}$ & $\sqrt{2}$ & $\sqrt{\frac{3d}{4}}$ \\
        \hline
    \end{tabular}
    \caption{Example of a simple table}
    \label{tab:example}
\end{table}

\rev{Let $\alpha(\ket{0}) = (1, 0, 0, \ldots, 0)^T$ be the vector representation of a $d$-dimensional quantum state $\ket{0}$, where the first coordinate is $1$ and the others are $0$. Let $\alpha(\ket{1}) = (0, 1, 0, \ldots, 0)^T$ be a $d$-dimensional vector with the second coordinate being $1$ and the others being $0$. 

Let $\alpha(\phi) = \frac{1}{\sqrt{d/2}}(1, \ldots, 1, 1, \ldots, 1, 0, \ldots, 0, 0, \ldots, 0)^T$ be the $d$-dimensional vector with the first $d/2$ coordinators being $1$ and second half being $0$, and $$\psi = \frac{1}{\sqrt{d/2}}(0, \ldots, 0, 1, \ldots, 1, 1, \ldots, 1, 0, \ldots, 0)^T$$ 
be the $d$-dimensional vector with the middle $d/2$ coordinates being $1$ and the rest being $0$.

Let $D(\phi, \psi)$ denote the trace distance between two quantum states $\phi$ and $\psi$. Let $L_1(\phi, \psi)$ and $L_2(\phi, \psi)$ denote the $\ell_1$ and $\ell_2$ distances between $\alpha(\phi)$ and $\alpha(\psi)$, respectively.  Let $L'_1(\phi, \psi)$ and $L'_2(\phi, \psi)$ denote the $\ell_1$ and $\ell_2$ distances between $p(\phi)$ and $p(\psi)$, where $p(\phi)$ is $\alpha(\phi)$ after taking the coordinate-wise absolute square; that is, $p(\phi) = \frac{1}{d/2}(1, \ldots, 1, 1, \ldots, 1, 0, \ldots, 0, 0, \ldots, 0)^T$.

The distortion of a distance $d(\cdot, \cdot)$ with respect to $D$ is lower bounded by
$$
    \sqrt{\max\bigg\{\frac{D(\phi, \psi)}{d(\phi, \psi)} \bigg/\frac{D(\ket{0}, \ket{1})}{d(\ket{0}, \ket{1})}, \frac{d(\phi, \psi)}{D(\phi, \psi)} \bigg/\frac{d(\ket{0}, \ket{1})}{D(\ket{0}, \ket{1})}\bigg\} }.
$$
In Table~\ref{tab:example}, we have calculated the distortions between the trace distance of the quantum states and the $\ell_1$ and $\ell_2$ distances of their corresponding classical vector representations.  It is easy to see that all distortions are larger than $1.1$\ .
}

\section{Search As A Special Case of Selection}
\label{app:search-as-selection}
We note that the selection operation can also be used for search. Given two quantum states ${\phi}$ and ${\psi}$, letting $M = \psi \psi^\dagger$, we have 
\begin{eqnarray*}
	D({\phi}, {\psi})  = \sqrt{1 - \abs{{\psi^\dagger}{\phi}}^2} = \sqrt{1 - \phi^\dagger M \phi}.
\end{eqnarray*}
Therefore, if we can estimate $\phi^\dagger M \phi$ up to an additive factor $c_{\veps}\veps^2$ (for a sufficiently small constant $c_{\veps}$) for any database state ${\phi}$, we can also solve $(\veps, \beta)$-search for any constant $\beta > 1$.   By Theorem~\ref{thm:QSS} and the fact that $\norm{M}_\infty^2 = 1$ when $M = \psi \psi^\dagger$, for a database consisting of $m$ states, the query (quantum) time using expectation value estimations is bounded by $9^n  m {\log m} \cdot \text{poly}(n) /{\veps^4} $. This approach is certainly much more time-expensive than that using state sketches presented in Section~\ref{sec:vector}.

%% file: norm.tex
\subsection{Proof of Theorem~\ref{thm:sketch-L1}}
\label{sec:proof-thm-sketch-L1}

\paragraph{Measurements Construction}
We first describe how to generate random measurements. We start by picking a random basis for $\CC^d$ based on the Haar measure, which can be done using a Gaussian ensemble of pure states as is used in \cite{HSW08}: Let $x_t, y_t\ (t = 1, \ldots, d)$ be independent Gaussian random variables with mean zero and variance $\sigma^2=\frac{1}{2d}$, and let  $g \triangleq (c_1, \ldots, c_d) \in \mathbb{C}^d$ be a random vector where $c_t=x_t + i y_t$. We repeat this process and generate $d$ Gaussian random vectors (written in the ket notation) $\ket{g_1}, \ldots, \ket{g_d}$.  

We next create an orthonormal basis for $\CC^d$ using $\ket{g_1}, \ldots, \ket{g_d}$. It is clear that with probability one, $\ket{g_t}$'s are linearly independent, which means that they span $\CC^d$.  However, they are not necessarily orthogonal.  To address this issue, we use the {\em pretty good measurement} technique \cite{Holevo13}. Define the operator  $\Gamma \triangleq \sum_{t \in [d]} \ketbra{g_t}$, and define the vector $\ket{\gamma_t} \triangleq \Gamma^{-1/2}\ket{g_t}$ for each $t \in [d]$.

We note that computing $\Gamma^{-1/2}$ may be time expensive. In Section B.1.1, we will discuss a more efficient measurement construction via {\em $t$-design}, which is a concept in quantum information theory that generalizes the idea of random sampling over the unitary group $U(d)$ of $d \times d$ unitary matrices.

\begin{claim}
The set $\left\{\ket{\gamma_t}: t \in [d]\right\}$ forms an orthonormal basis for $\CC^d$.
\end{claim}
\begin{proof}
Observe that
\begin{equation*}
\sum_{t \in [d]} \ketbra{\gamma_t} = \Gamma^{-1/2}\left( \sum_{t \in [d]}\ketbra{g_t}\right)\Gamma^{-1/2} = I. 
\end{equation*}
Moreover, $\ket{\gamma_t}$ are linearly independent since $\ket{g_t}$'s are linearly independent.  Hence, $\ket{\gamma_t}$'s are orthonormal. 
\end{proof}
We also note that the distribution of $\ket{\gamma_t}$ is unitary invariant. Hence, it is the Haar measure. 

Next, we randomly group $\ket{\gamma_t}$'s into $k$ groups and form random projection operators as 
\begin{equation}
\Pi_j = \sum_{\ell \in [d/k]} \ketbra{\gamma^{j}_\ell}, \qquad j=1, \ldots, k.
\label{eq:Pi}
\end{equation}
Let $\mathcal{M}_k = \{\Pi_1, \cdots, \Pi_k\}$ be the corresponding measurement. Clearly, $\mathcal{M}$ is a valid POVM with probability one. 

\paragraph{The Analysis of Distortion} Let $\M_k(\ket{\phi})$ and  $\M_k(\ket{\psi})$ be the probability distributions of the measurement outcomes when the states are $\ket{\phi}$ and $\ket{\psi}$, respectively. Then, the total variation distance between the two probability distributions can be written as  
\begin{equation}
\norm{\M_k(\ket{\phi}) - \M_k(\ket{\psi})}_1  = \sum_{j \in [k]} \abs{\tr{\Pi_j \ketbra{\phi}} - \tr{\Pi_j \ketbra{\psi}}}
= \sum_{j \in [k]} \abs{\tr{\Pi_j A}}, \label{eq:1norm-a}
\end{equation}
where we have used the Born's law and set $A \triangleq \ketbra{\phi}-\ketbra{\psi}$. 

We next show that $A$ has two eigenvalues $\pm D(\ket{\phi}, \ket{\psi})$, where $D(\cdot, \cdot)$ is the trace distance. 
To this end, suppose $\ket{\omega}$ is an eigenstate of $A$, and $A\ket{\omega}=\lambda \ket{\omega}$, where $\lambda\in \RR$ as $A$ is a Hermitian operator. Multiplying both sides by $\bra{\phi}$ gives 
\begin{equation}
	\label{eq:e-1}
\matrixel{\phi}{A}{\omega}= \braket{\phi}{\omega} - \braket{\phi}{\psi}\braket{\psi}{\omega} = \lambda \braket{\phi}{\omega}.
\end{equation}
Similarly, multiplying both sides by $\bra{\psi}$ gives 
\begin{equation}
		\label{eq:e-2}
\matrixel{\psi}{A}{\omega}= \braket{\psi}{\phi} \braket{\phi}{\omega} - \braket{\psi}{\omega} = \lambda \braket{\psi}{\omega}. 
\end{equation}
Combining (\ref{eq:e-1}) and (\ref{eq:e-2}) gives
\begin{align*}
(1+\lambda) \braket{\psi}{\omega}= \braket{\psi}{\phi} \braket{\phi}{\omega} = \braket{\psi}{\phi} \frac{1}{1-\lambda} \braket{\phi}{\psi}\braket{\psi}{\omega}.
\end{align*}
Hence, $(1-\lambda)(1+\lambda) = \abs{\braket{\phi}{\psi}}^2$, which implies 
\begin{equation}
	\label{eq:f-1}
	\lambda = \pm \sqrt{1-\abs{\braket{\phi}{\psi}}^2 } = \pm D(\ket{\phi}, \ket{\psi}). 
\end{equation}
Now without loss of generality, let $\ket{1}$ and $\ket{2}$ denote the two eigenstates of $A$. Hence, 
\begin{equation}
\label{eq:f-11}
A= \abs{\lambda}(\ketbra{1}-\ketbra{2}).
\end{equation} 
The right-hand side of \eqref{eq:1norm-a} can be written as
\begin{eqnarray}
\sum_{j \in [k]} \abs{\tr{\Pi_j A}} &=& \abs{\lambda} \sum_{j \in [k]} \abs{  \expval{\Pi_j}{1} -\expval{\Pi_j}{2}  } \\
&=& \abs{\lambda} \sum_{j \in [k]} \abs{  \sum_{\ell \in [d/k]} \abs{\braket{1}{\gamma^{j}_\ell}}^2 -\abs{\braket{2}{\gamma^{j}_\ell}}^2  }. \label{eq:f-12}
\end{eqnarray}
Let $W_\ell^{j} \triangleq d \left(\abs{\braket{1}{\gamma^{j}_\ell}}^2 -\abs{\braket{2}{\gamma^{j}_\ell}}^2\right)$. Combining \eqref{eq:1norm-a} and \eqref{eq:f-12}, we have
\begin{equation*}
\norm{\M_k(\ket{\phi}) - \M_k(\ket{\psi})}_1  =\abs{\lambda} \sum_{j \in [k]} \abs{\sum_{\ell \in [d/k]} \frac{1}{d}W_\ell^{j}}.
\end{equation*}
Multiplying both sides of the above equality by $\sqrt{\frac{d}{k}}$ gives 
\begin{eqnarray}
\sqrt{\frac{d}{k}} \norm{\M_k(\ket{\phi}) - \M_k(\ket{\psi})}_1 
&=&  \frac{\abs{\lambda}}{k}\sum_{j \in [k]} \abs{ \frac{1}{\sqrt{d/k}} \sum_{\ell \in [d/k]} W_\ell^{j}  } \nonumber \\
&\stackrel{\eqref{eq:f-1}}{=}&  \frac{D(\ket{\phi}, \ket{\psi})}{k}\sum_{j \in [k]} \abs{ \frac{1}{\sqrt{d/k}} \sum_{\ell \in [d/k]} W_\ell^{j}  }.
	\label{eq:norm-1b}
\end{eqnarray}

We try to analyze the expectation and variance of each $W_\ell^{j}$.
First, note that $\EE\left[W_\ell^{j}\right] = 0$, because the distribution of $\ket{\gamma^{j}_\ell}$ is unitary invariant, which implies that $\abs{\braket{1}{\gamma^{j}_\ell}}^2$ and  $\abs{\braket{2}{\gamma^{j}_\ell}}^2$ have an identical distribution. 

The variance of $W_\ell^{j}$ equals to
\begin{eqnarray}
\sigma_W^2 &\triangleq& \mathbf{Var}\left[W_\ell^{j}\right] = \EE\left[|W_\ell^{j}|^2\right] \nonumber\\
&=&{d^2}\EE\left[\left(\abs{\braket{1}{\gamma^{j}_\ell}}^2 -\abs{\braket{2}{\gamma^{j}_\ell}}^2\right)^2 \right] \nonumber\\
&=&{2}{d^2} \left( \EE\left[\abs{\braket{1}{\gamma}}^4\right]-\left(\EE\left[\abs{\braket{1}{\gamma}}^2 \abs{\braket{2}{\gamma}}^2\right]\right)\right),\quad \label{eq:g-1}
\end{eqnarray}
where $\ket{\gamma}$ is a random pure state generated based on the Haar measure. Since the Haar measure is invariant under Unitary transformation,  $\braket{1}{\gamma}$ and $\braket{2}{\gamma}$ have the same joint distribution as $U_{11}$ and $U_{21}$, where $U$ is a random unitary matrix (a Haar unitary) and $U_{ij}$ refers to the entry on the $i$ths row and $j$th column of $U$. Note that all entries $U_{ij}$ of a Haar unitary $U$ are identically distributed \cite{HP06}. Moreover, they can be written as $U_{ij}=re^{i\theta}$ with the distribution given by  $\frac{d-1}{\pi}(1-r^2)^{d-2}r\partial r\partial\theta$ where $r\in [0,1]$ and $\theta\in [0,2\pi]$. Therefore, the distribution of $|U_{11}|^2$  is given by $(d-1)(1-r)^{d-2}\partial r$. Consequently, from (\ref{eq:g-1}) we have 
\begin{equation}
	\label{eq:g-2}
\sigma_W^2 ={2}{d^2} \left(\EE\left[|U_{11}|^4\right]-\EE\left[|U_{11}|^2|U_{21}|^2\right]\right).
\end{equation}
The first expectation in (\ref{eq:g-2}) calculates as 
\begin{equation*}
\EE[|U_{11}|^4] = (d-1)\int_{0}^{1}r^2(1-r)^{d-2}\partial r = (d-1)\mathcal{B}(3, d-1),
\end{equation*}
where $\mathcal{B}(\cdot, \cdot)$ is the Beta function that is defined as 
\begin{equation*}
\mathcal{B}(\alpha, \beta)= \int_0^1 r^{\alpha-1}(1-r)^{\beta-1} \partial r.
\end{equation*}
The Beta function at positive integers can be calculated combinatorically as 
\begin{equation*}
\mathcal{B}(m,n) =\frac{(m + n)/(mn)}{\binom{m + n}{m}}.
\end{equation*}
Therefore, 
\begin{equation*}
\EE\left[|U_{11}|^4\right] = (d-1)  \frac{(d+2)/ (3(d-1))}{\binom{d+2}{3}} = \frac{2}{d(d+1)}.
\end{equation*}
Similarly, the second expectation in (\ref{eq:g-2}) equals to 
\begin{equation*}
\EE\left[|U_{11}|^2|U_{21}|^2\right] = \frac{1}{d(d+1)}.
\end{equation*}
Consequently, (\ref{eq:g-2}) reduces to
\begin{equation}
	\label{eq:g-3}
\sigma_W^2 = {2}{d^2} \left(\frac{2}{d(d+1)} - \frac{1}{d(d+1)}\right) = \frac{2d}{d+1}.
\end{equation}
Implying that $\sigma_W^2\leq 2$.
Now we continue our investigation of Equality~(\ref{eq:norm-1b}). Let 
\begin{equation*}
	\label{eq:f-2}
	Z^{j} \triangleq  \frac{1}{\sqrt{d/k}} \sum_{\ell \in [d/k]} W_\ell^{j}. 
\end{equation*}
\eqref{eq:norm-1b}  simplifies to  
\begin{equation}
	\label{eq:norm-c}
Q \triangleq \sqrt{\frac{d}{k}} \norm{\M_k(\ket{\phi}) - \M_k(\ket{\psi})}_1  = D(\ket{\phi}, \ket{\psi}) \cdot \frac{1}{k}\sum_{j \in [k]} |Z^{j}|.
\end{equation}
Let 
\begin{equation}
	\label{eq:f-3}
U \triangleq \frac{Q - \EE[Q]}{D(\ket{\phi}, \ket{\psi})}
= \frac{1}{k}\sum_{j \in [k]}  \left( |Z^{j}| - \EE\left[|Z^{j}|\right]\right).
\end{equation}

Applying the generic Chernoff bound (Lemma~\ref{lem:chernoff}) to $U$ gives
\begin{equation}\label{eq:chernoff}
\PP[U \geq \eta] \leq \inf_{t>0}M_{U}(t)e^{-t\eta},
\end{equation}
where $M_{U}(t) = \EE\left[e^{t U}\right]$ is the moment generating function.

Let $t_{0} \triangleq \eta k/2$. We use $t = t_{0}$ to upper bound the RHS of (\ref{eq:chernoff}). 
Since $Z^{j}$'s are i.i.d., we have 
\begin{equation}
M_{U}(t_{0}) = \prod_{j \in [k]} \EE\left[\exp\left({\frac{t_{0}}{k}\left( |Z^{j}| - \EE\left[|Z^{j}|\right]\right)}\right)\right] =\left( M_{\tilde{U}}\left(\frac{t_{0}}{k}\right)\right)^k = \left( M_{\tilde{U}}\left(\frac{\eta}{2}\right)\right)^k,
\label{eq:h-1}
\end{equation}
where $\tilde{U}: = |Z^{j}| - \EE\left[|Z^{j}|\right]$.
Apply Taylor's expansion on $M_{\tilde{U}}\left(\frac{\eta}{2}\right)$ around $\eta=0$, we have 
\begin{equation}
M_{\tilde{U}}\left(\frac{\eta}{2}\right) = M_{\tilde{U}}(0) + M'_{\tilde{U}}(0) \cdot \frac{\eta}{2}+M^{''}_{\tilde{U}}(0) \cdot \frac{1}{2}\left(\frac{\eta}{2}\right)^2+\zeta_{\eta},
\label{eq:h-2}
\end{equation}
where $\zeta_{\eta} \le c_\eta \eta^3$ (for a universal constant $c_\eta > 0$)  is the remainder term.

It is clear that $M_{\tilde{U}}(0) = 1$. By the definition of the moment generating function, $M'_{\tilde{U}}(0) = \EE\left[\tilde{U}\right]=0$ and 
\begin{equation*}
M^{''}_{\tilde{U}}(0) = \mathbf{Var}\left[\tilde{U}\right] = \mathbf{Var}\left[|Z^{j}|\right] \le \mathbf{Var}\left[Z^{j}\right] = \mathbf{Var}\left[W_\ell^{j}\right]  = \sigma^2_W\ ,
\end{equation*}
(\ref{eq:h-2}) simplifies to the following: 
\begin{equation*}
M_{\tilde{U}}\left(\frac{\eta}{2}\right) \le 1 + \frac{\sigma^2_W }{2}\left(\frac{\eta}{2}\right)^2+c_\eta \eta^3 \le 1 + \left(\frac{\eta}{2}\right)^2+c_\eta \eta^3,
\end{equation*}
where we have used the fact that $\sigma_W^2 \le 2$ (see (\ref{eq:g-3})).

In the rest of the analysis, we will focus on the parameter $\eta \le \frac{1}{8c_\eta}$; the actual value of $\eta$ will be determined later.

Hence, by (\ref{eq:h-1}) we have
\begin{equation}
M_{U}(t_0) \le \left(1 + \left(\frac{\eta}{2}\right)^2+c_\eta \eta^3\right)^k
\le \exp\left(\frac{\eta^2 k}{4}+ c_\eta \eta^3 k\right),
\label{eq:i-2}
\end{equation}
where in the second inequality we have used the fact 
$\left(1 + \frac{x}{k}\right)^k \le e^x$. 

Plugging (\ref{eq:i-2}) to (\ref{eq:chernoff}), we have
\begin{equation*}
	\PP(U \geq \eta ) 
	\le \exp\left(\frac{\eta^2 k}{4}+ c_\eta \eta^3 k \right) \cdot \exp\left(- \frac{\eta^2 k}{2} \right)  = \exp\left(-\Omega(\eta^2 k)\right). 
\end{equation*}
where we have used $c_\eta \eta^3 k \le \frac{\eta^2 k}{8}$ since $\eta \le \frac{1}{8c_\eta}$.

For the other direction, we have
\begin{equation}
\label{eq:i-21}
	\PP\left[U \le -\eta \right] = \PP \left[- U \ge \eta \right] \le M_U (-t_0) e^{-t_0 \eta} \le \exp\left(-\Omega(\eta^2 k)\right). 
\end{equation}
Hence, for any $\delta\in (0,1)$, setting  $k= c_k \left(\frac{1}{\eta^2}\log \frac{1}{\delta}\right)$ for a  sufficiently large constant $c_k > 0$, we have that $|U| \leq \eta$ with probability at least $(1-\delta)$.  Combining \eqref{eq:i-21},  (\ref{eq:f-3}),  and \eqref{eq:norm-c}, we have that with probability at least $(1-\delta)$,
\begin{equation}
	\label{eq:i-3}
	\abs{\sqrt{\frac{d}{k}} \cdot \frac{\norm{\M_k(\ket{\phi}) - \M_k(\ket{\psi})}_1}{D(\ket{\phi}, \ket{\psi})}   - \EE[\abs{Z}]}\leq \eta, 
\end{equation}
where $Z$ is distributed identically as $Z^j$'s.

It remains to investigate $\EE[|Z|]$.  We show that $\EE[|Z|] = \Theta(1)$.  By the definition we know that 
\begin{equation*}
\EE[Z] = \frac{1}{\sqrt{d/k}} \sum_{\ell \in [d/k]} \EE[W_\ell^1] = 0.
\end{equation*}
We start with the upper bound.
\begin{equation}
\EE[|Z|] \le \sqrt{\EE[|Z|^2]} = \sqrt{\EE[Z^2]} = \sqrt{\mathbf{Var}[Z]} = \sigma_W \le \sqrt{2}.
\label{eq:i-5}
\end{equation}

For the lower bound, Markov inequality implies that 
\begin{align*}
\EE\left[|Z|\right] \geq a \PP\left[|Z| >a\right],
\end{align*}
for any $a>0$.
Since $Z$ is distributed identically as sum of i.i.d.\ random variables $\{W_\ell^1\}_{\ell \in [d/k]}$ with $E[W_\ell^1] = 0$, $E[(W_\ell^1)^2] = \sigma_W^2$, and $E[(W_\ell^1)^3] < +\infty$, by the Berry-Esseen theorem (Lemma~\ref{lem:berry-esseen})  we get
\begin{equation*}
\PP\left[\frac{1}{\sigma_W}|Z| > x\right] 
 \geq \PP[|N(0,1)| \geq x] -O\left(\frac{1}{\sqrt{d/k}}\right) = \left(1-\erf\left(\frac{x}{\sqrt{2}}\right)\right) - O\left(\frac{1}{\sqrt{d/k}}\right).
\end{equation*}
Hence, 
\begin{equation}
	\label{eq:j-3}
\EE\left[|Z|\right] \geq \sup_{x>0} \sigma_W x  \left(1-\erf\left(\frac{x}{\sqrt{2}}\right)\right) - O\left(\frac{1}{\sqrt{d/k}}\right).
\end{equation}
Recall that $\sigma_W\approx \sqrt{2}$ (see (\ref{eq:g-3})). By a numerical computation, the maximum of the RHS of (\ref{eq:j-3}) attends at $x\approx 0.7475$, which gives 
\begin{equation}
	 \label{eq:j-2}
\EE\left[|Z|\right] \geq 0.4807 - O\left(\frac{1}{\sqrt{d/k}}\right) \ge 0.48
,
\end{equation}
as long as $d/k$ is a sufficiently large constant.

By (\ref{eq:i-5}) and (\ref{eq:j-2}), we have $\mu_Z \triangleq \EE[\abs{Z}] = \Theta(1)$.  Now, for any given constant $\iota > 0$, we set the $\eta = \min\left\{\iota \mu_Z, \frac{1}{8c_\eta}\right\}$. (\ref{eq:i-3}) gives
\begin{equation*}
	\label{eq:i-4}
	\abs{\sqrt{\frac{d}{k}} \frac{1}{\mu_Z}  \frac{\norm{\M_k(\ket{\phi}) - \M_k(\ket{\psi})}_1}{D(\ket{\phi}, \ket{\psi})}  - 1}\leq \iota .
\end{equation*}

\subsubsection{Efficient Implementations of Random Measurements} 
\label{sec:efficient-impl}
Lastly, we address the space and time complexity of the sketching procedure in this proof. Specifically, we consider efficient construction of $\mathcal{M}_k$. Note that the runtime of the original measurement construction based on pretty good measurements is polynomial in $d$, hence exponential in the number of qubits. This is because it relies on generating $d$ kets $\ket{\gamma_\ell^j}$ based on Haar measure and via complex Gaussian vectors in $\CC^d$ and the inverse-square root of the matrix $\Gamma$. Generally, the classical time and the circuit gate complexity for sampling from Haar distribution is exponential. In what follows, we present a construction for $\mathcal{M}_k$ with $\text{poly}\log(d)$ time and space.

Our approach is based on 2-design methods. A unitary $t$-design is a concept in quantum information theory that generalizes the idea of random sampling over the unitary group $U(d)$ of $d \times d$ unitary matrices. It provides a way to approximate certain statistical properties of quantum states or operations without needing to sample from the entire group, which can be computationally expensive. Below we highlight basics of this concept.

Let $P_{t,t}(U)$ denote  a polynomial  which is homogeneous with degree at most $t$ in the matrix elements of $U$, and at most degree $t$ in the complex conjugates of these elements. 
\begin{definition}
A unitary $t$-design is a finite set of unitary matrices $\{U^{(i)}\}_{i=1}^N$ such that for any homogeneous polynomial $P_{t,t}$
\[
\frac{1}{N} \sum_{i=1}^{N} P_{t,t}(U^{(i)}) = \int P_{t,t} \, d\mu_{\text{Haar}}(U)
\]
where $\mu_{\text{Haar}}$ is the Haar measure on the space of $d \times d$ unitary matrices.
\end{definition}
Intuitively, this definition implies that a unitary $t$-design is indistinguishable  from Haar measure when only polynomials of degree at most $t$ are used. We show that a 2-design is sufficient to obtain Theorem \ref{thm:sketch-L1}. 

Note that the analysis of sample complexity in our proof relies on the generic Chernoff bound which, per \eqref{eq:h-2}, depends on the first two moments of $$W_\ell^j = d \left(\abs{\braket{1}{\gamma_\ell^j}}^2 -\abs{\braket{2}{\gamma_\ell^j}}^2\right).$$ 
We show that $W_\ell^j$ can be written as a polynomial $P_{1,1}$ of degree $t=1$ in the elements of a unitary matrix $U$. Let $U$ be a matrix with columns being the vectors of $\ket{\gamma_\ell^j}$. That is $\braket{r}{\gamma_\ell^j}$ is the element of $U$, denoted by $U_{r, (\ell, j)}$, located at the row $r$ and the column indexed by $(\ell, j)$.  With this definition, we can write   $\abs{\braket{1}{\gamma_\ell^j}}^2 = U_{1, (\ell, j)} U_{1, (\ell, j)}^*$, implying that  $W_\ell^j = P_{1,1}(U)$. 
Therefore, the first moment of $W_\ell^j$ is a polynomial $P_{1,1}$ of degree $t=1$ in $U$. Moreover, the second moment of $W_\ell^j$ is a   polynomial $P_{2,2}$ of degree $t=2$ in $U$. This is because based on \eqref{eq:g-1}, the second moment is a function of $\abs{\braket{1}{\gamma_\ell^j}}^4$, which is written as $U_{1, (\ell, j)}^2 (U_{1, (\ell, j)}^*)^2$. Hence, as long as the first and the second moments of $W_\ell^j$ resemble the Haar measure, the proof remains valid, indicating that a 2-design is sufficient. 

Based on the above argument, instead of using the pretty good measurement construction, we sample from $2$-design unitaries. This can be implemented efficiently using the Clifford group, a specialized class of quantum operators (for more details see Section \ref{sec:clifford}). It is well-known that the Clifford group is a 2-design \cite{Dankert2009}. Furthermore, there exists an algorithm that samples uniformly from the Clifford group in classical time $O(n^8)$ and outputs a circuit representing the measurement with a gate complexity of $O(n^2)$, where $n = \log d$ is the number of qubits \cite{DiVincenzo2002}. To construct the desired measurement $\mathcal{M}_k$, we first randomly generate a Clifford circuit, apply it to the input quantum state, and then measure in the computational basis. The outcome of the measurement is a binary string in $\{0,1\}^n$. Lastly, we design a binning function $f$ that takes the $n$-bit string and outputs an index in $[k]$.
This is achieved by partitioning the set $\{0,1\}^n$ into $k$ equal-size bins, which can be efficiently implemented using a decision tree that only reads the first $\log k$ bits of the binary measurement outcomes. There are $k$ possible outputs of the decision tree, meaning that it can be represented by a function $f: \{0,1\}^n\rightarrow [k]$ that determines the bin index for any binary string of length $n$. This construction is summarized as Algorithm \ref{alg:Mk}.  
\begin{algorithm}[t]
\caption{$k$-sketching measurement}
\label{alg:Mk}
\DontPrintSemicolon
\SetKwComment{Comment}{/* }{ */}
\KwIn{$k, n, \ket{\phi}$}
 \SetKwProg{Fn}{Function}{:}{} 
Sample from the Clifford group on $n$-qubits and construct the random circuit $U$.\;
Apply $U$ on the input state $\ket{\phi}$\;
Measure the first $\lceil\log k\rceil$ qubits along the computational basis.\;
\KwRet mod-$k$ of the decimal representation of the resulted binary string.\;   
\end{algorithm}

In conclusion, the above construction generates a random measurement $\mathcal{M}_k$ with $O(n^2)$ quantum gates  and $O(n^8)$ classical time. Moreover, this measurement  enjoys the same bound on the sample complexity as for the original construction based on the Haar measure.  

\subsection{Proof of Theorem~\ref{thm:sketch-L2}}
\label{sec:proof-thm-sketch-L2}

The proof for Theorem~\ref{thm:sketch-L2} (the $\ell_2$-norm case) is similar to that for Theorem~\ref{thm:sketch-L1} (the $\ell_1$-norm case), but the calculation will be different due to different distance functions.  Let $\M_k = \{\Pi_1, \ldots, \Pi_k\}$ be the same random POVM generated as that in the proof of Theorem~\ref{thm:sketch-L1}. 

The $\ell_2$ distance between the output probability vectors can be written as
\begin{equation}
\norm{\M_k(\ket{\phi}) - \M_k(\ket{\psi})}^2_2  = \sum_{j \in [k]} \abs{\tr{\Pi_j \ketbra{\phi}} - \tr{\Pi_j \ketbra{\psi}} }^2 = \sum_{j \in [k]} \abs{\tr{\Pi_j A}}^2, \label{eq:2norm-a}
\end{equation}
where $A=\ketbra{\phi}-\ketbra{\psi}$, which can be rewritten as $A=|\lambda| (\ketbra{1}-\ketbra{2})$, where $\ket{1}$ and $\ket{2}$ are two eigenstates of $A$. Hence, we can rewrite (\ref{eq:2norm-a}) as
\begin{eqnarray}
 \norm{\M_k(\ket{\phi}) - \M_k(\ket{\psi})}^2_2 \nonumber 
&=& \sum_{j \in [k]} \abs{\tr{\Pi_j A}}^2 \nonumber \\
&=&
\abs{\lambda}^2 \sum_{j \in [k]} \abs{\expval{\Pi_j}{1} -\expval{\Pi_j}{2}  }^2 \nonumber \\
&=& \abs{\lambda}^2 \sum_{j \in [k]} \left(\sum_{\ell \in [d/k]} \abs{\braket{1}{\gamma^{(j)}_\ell}}^2 -\abs{\braket{2}{\gamma^{(j)}_\ell}}^2  \right)^2.
\label{eq:k-1}
\end{eqnarray}
Multiplying both sides of (\ref{eq:k-1}) by a factor of $d$, and letting $W_\ell^{j} \triangleq d \left(\abs{\braket{1}{\gamma^{j}_\ell}}^2 -\abs{\braket{2}{\gamma^{j}_\ell}}^2\right)$,  we have
\begin{equation}
	d \norm{\M_k(\ket{\phi}) - \M_k(\ket{\psi})}^2_2 = \abs{\lambda}^2 \frac{1}{k} \sum_{j \in [k]} \left(\frac{1}{\sqrt{d/k}} \sum_{\ell \in [d/k]} W_\ell^j \right)^2. \label{eq:k-2}
\end{equation}
Let 
\begin{equation}
	\label{eq:k-4}
Z^j \triangleq \frac{1}{\sqrt{d/k}} \sum_{\ell \in [d/k]} W_\ell^j.
\end{equation}
By a similar analysis as that in the proof of Theorem~\ref{thm:sketch-L1} (particularly, recall (\ref{eq:f-1}) and (\ref{eq:g-3})), the expectation of the RHS of (\ref{eq:k-2}) is calculated to be 
\begin{equation}
\label{eq:k-3}
\EE\left[\abs{\lambda}^2 \frac{1}{k} \sum_{j \in [k]} |Z^j|^2\right] =  \abs{\lambda}^2 \sigma_W^2 = D^2(\ket{\phi}, \ket{\psi})\cdot \frac{2d}{d+1}.
\end{equation} 
We next analyze the variance of the RHS of (\ref{eq:k-2}).

We write
\begin{equation}
V \triangleq \frac{1}{|\lambda|^2}\left(d \norm{\M_k(\ket{\phi}) - \M_k(\ket{\psi})}_2^2 - |\lambda|^2 \sigma_W^2 \right)
= \frac{1}{k}\sum_{j \in [k]}  \left(|Z^j|^2 -  \sigma_W^2\right). \label{eq:l-0}
\end{equation}
By the generic Chernoff bound (Lemma~\ref{lem:chernoff}), we have
\begin{equation}
\label{eq:l-1}
\PP(V \geq \eta ) \leq \inf_{t>0}M_{V}(t)e^{-t \eta},
\end{equation}
where $M_V(t) = \EE\left[e^{t V}\right]$.  Set $t_0 = c_t \eta k$ for a constant $c_t$ to be determined later .  We use $t = t_0$ to upper bound the RHS of (\ref{eq:l-1}). Since $Z^j$'s are i.i.d., we have
\begin{equation}
M_V(t_0) = \prod_{j \in [k]} \EE\left[\exp\left({\frac{t_{0}}{k}\left( |Z^{j}|^2 - \sigma_W^2\right)}\right)\right] = \left(M_{\tilde{V}}\left(\frac{t_0}{k}\right)\right)^k = \left(M_{\tilde{V}}\left(c_t \eta\right) \right)^k,
\label{eq:l-2}
\end{equation}
where $\tilde{V} = |Z^{j}|^2 - \sigma_W^2$.  

Apply Taylor's expansion on $M_{\tilde{V}}\left(c_t \eta\right)$ around $\eta = 0$, we get
\begin{equation}
	M_{\tilde{V}}\left(c_t \eta\right) = M_{\tilde{V}}(0) + M'_{\tilde{V}}(0) \cdot c_t \eta+M^{''}_{\tilde{V}}(0) \cdot \frac{1}{2}\left(c_t \eta\right)^2+\zeta_{\eta},
	\label{eq:l-3}
\end{equation}
where $\zeta_{\eta} \le c_\eta \eta^3$ (for a universal constant $c_\eta > 0$) is the remainder term.  We again have $M_{\tilde{V}}(0) = 1$, $M'_{\tilde{V}}(0) = \EE\left[\tilde{V}\right] = 0$, and $M^{''}_{\tilde{V}}(0) = \mathbf{Var}\left[\tilde{V}\right] \le c_{\tilde{V}}$ for a universal constant $c_{\tilde{V}} > 0$.

In the rest of the analysis, we will focus on parameter
$\eta  \le \frac{c_{\tilde{V}} c_t^2}{2c_\eta}$; the actual value of $\eta$ will be determined later.

We extend (\ref{eq:l-3}) as
\begin{equation}
	M_{\tilde{V}}\left(c_t \eta\right) \le 1 + \frac{c_{\tilde{V}} }{2}\left(c_t \eta\right)^2+c_\eta \eta^3 \le 1 + c_{\tilde{V}} c_t^2 \eta^2 .  	\label{eq:l-4}
\end{equation}
Plugging (\ref{eq:l-4}) to (\ref{eq:l-2}), we have
\begin{equation}
	M_V(t_0) =  \left(M_{\tilde{V}}\left(c_t \eta\right)\right)^k \le  \left(1 + c_{\tilde{V}} c_t^2 \eta^2 \right)^k 
    \le \exp\left(c_{\tilde{V}} c_t^2 k \eta^2\right) . \label{eq:m-1}
\end{equation}
Plugging (\ref{eq:m-1}) to (\ref{eq:l-1}), we have 
\begin{equation}
	\PP(V \ge \eta) \le M_V(t_0) e^{-t_0 \eta} 
 \le \exp\left(c_t^2 c_{\tilde{V}} k \eta^2\right) \exp\left(-c_t \eta^2 k\right)
 = \exp\left(-\Omega(\eta^2 k)\right), \nonumber
\end{equation}
where for the last equality to hold, we set constant $c_t = \frac{1}{2c_{\tilde{V}}}$.

For the other direction, we have
\begin{equation}
	\PP\left[V \le -\eta \right] = \PP \left[- V \ge \eta \right] \le M_V(-t_0) e^{-t_0 \eta} \le \exp\left(-\Omega(\eta^2 k)\right). \nonumber
\end{equation}

Hence, for any $\delta \in (0, 1)$, setting $k = c_k \left(\frac{1}{\eta^2}\log \frac{1}{\delta}\right)$ for a  sufficiently large constant $c_k > 0$, we have that $|V| \leq \eta$ with probability at least $(1-\delta)$.  This, together with (\ref{eq:k-2}), (\ref{eq:k-4}) (\ref{eq:k-3}), and (\ref{eq:l-0}), we obtain
\begin{equation*}
	\abs{\frac{d \norm{\M_k(\ket{\phi})- \M_k(\ket{\psi})}^2_2}{D^2(\ket{\phi}, \ket{\psi})}  - \frac{2d}{d+1}} \le \eta ,
\end{equation*}
which implies
\begin{equation}
	\label{eq:m-2}
	 \sqrt{2 - \eta - o(1)} \le \frac{\sqrt{d} \norm{\M_k(\ket{\phi})- \M_k(\ket{\psi})}_2}{D(\ket{\phi}, \ket{\psi})} \le \sqrt{2 + \eta}.
\end{equation}
Now, for any constant $\iota > 0$, we set $\eta  = \left\{\iota, \frac{c_{\tilde{V}} c_t^2}{2c_\eta} \right\}$. (\ref{eq:m-2}) gives
\begin{equation*}
	\abs{\sqrt{\frac{d}{2}} \cdot \frac{\norm{\M_k(\ket{\phi})- \M_k(\ket{\psi})}_2}{D(\ket{\phi}, \ket{\psi})} - 1} \le \iota .
\end{equation*}

%% file: QSS.tex
\subsection{Proof of Theorem~\ref{thm:QSS}}
\label{app:QSS}
\rev{For the first part, we use the same procedure is CST to create the $N \times n$ seed matrix $A({\phi})$, where each row $i\in [N]$ consists of the $n$ pairs $\{b_{i,j}, \mathtt{index}(U_{i,j}))\}_{j = 1}^n$.}

\rev{Next, for the query algorithm, we propose an encoding to turn the classical shadows into quantum states. This is a deviation from CST as we push the computations back to quantum via a QCQC approach. Let $Q$ be the index of the relevant qubits of a local observable $M$ in the query phase.  For each $j \in Q$, we create a random binary pair $(c_{i,j}, w_{i,j})$ using the stored bit $b_{i,j}$ as follows:
\begin{equation}\label{eq:seed}
	(c_{i,j}, w_{i,j}) = 
	\begin{cases} 
		(b_{i,j}, 1) & \text{w.pr.\ } 2/3\ , \\
		(1-b_{i,j}, -1) & \text{w.pr.\ } 1/3\ . 
	\end{cases}
\end{equation}
We then prepare a qubit $\ket{c_{i, j}}$ and apply the corresponding operator $U_{i, j}$ to create qubit $\ket{v_{i, j}} = U_{i, j} \ket{c_{i, j}}$. Note that  $\ket{v_{i, j}}$ takes one of the following states:
\[
\ket{0}, \ket{1}, \ket{+}, \ket{-}, \ket{+i}, \ket{-i},
\]
that are easy to prepare (see Appendix~\ref{app:gate} for their vector representations). Then, we prepare the $\ket{0}$ for the rest of qubits not indexed in $Q$. Let
\begin{equation}
	\label{eq:p-1}
	\ket{v'_{i,j}} = 
	\begin{cases} 
		\ket{v_{i,j}}, & \text{if\ } j \in Q \\
		\ket{0}, & \text{otherwise}.
	\end{cases}
\end{equation}
We construct the $i$-th shadow sample as (written as the outer-product form)
\begin{equation}
	\label{eq:p-11}
	 \ket{\tilde{\phi}_i}= \bigotimes_{j = 1}^n \ket{v'_{i, j}}.
\end{equation}
We next measure each $\ket{\tilde{\phi}_i}$ using the observable $M$, and obtain an outcome $x_i$.  Let
\begin{equation}
	\label{eq:p-2}
	S_i = 3^k x_i \prod_{j \in Q} w_{i, j}.
\end{equation}
Then, our estimator is the empirical average 
\begin{equation*}
T=\frac{1}{N} \sum_i S_i.
\end{equation*}

We now show that when $N \ge 9^{k} \norm{M}_\infty^2  \frac{\log(1/\delta)}{\veps^2}$, then $T$ approximates $\ev{M}{\phi}$ up to an additive error $\veps$ with probability $(1-\delta)$, proving the correctness part of Theorem~\ref{thm:QSS}.  We will also give the query time analysis.  Recall that the space needed for storing the seed matrix is $O(N n)$ classical bits.
}

In the rest of the proof, we will focus on the random variable $S \triangleq S_i$ for particular $i \in [N]$.  Recall that the final approximation $T$ is the average of $N$ i.i.d.\ copies of $S$.  

Let $X_i$, $W_j$, $V_{i,j}$, $B_{i,j}$ be the corresponding random variables of $x_i$, $w_{i,j}$, $v_{i,j}$, $b_{i,j}$, respectively.  Since we focus on a particular $i \in [N]$, we will omit all subscripts $i$ in  those random variables and write them as  $X$, $W_j$, $V_j$, $B_j$.  We also write $\ket{\tilde{\phi}_i}$ as $\ket{\tilde{\phi}}$.

The following result can be inferred from \cite{HNX+23}.  We include a proof for completeness.

\begin{lemma}
	For any $i \in [N]$, let $\ketbra{\tilde{\varphi}_i}  = \bigotimes_{j = 1}^n \ketbra{V_j}$.  We have $\EE\left[3^n W \ketbra{\tilde{\varphi}_i}\right] = \ketbra{\phi}$, where $W = \prod_{j=1}^n W_j$.
\end{lemma}

\begin{proof}

\rev{Let \(\Gamma_0\) denote the shadow channel for Pauli measurements as defined in \cite{HKP20}, which is given by}
 	\begin{equation*}
		\Gamma_0[O] \triangleq \sum_{U\in \{ I, H, S^\dagger H\}}\sum_{b\in \{0,1\}} \frac{1}{3}\matrixelement{b}{U^{\dagger} O U}{b}~ U\ketbra{b}U^\dagger,
	\end{equation*} 
 for any single qubit operator $O$.
	By direct calculation, we have for a generic state $\ket{\psi}=a_0 \ket{0} + a_1 \ket{1}$, 
	\begin{equation*}
		\Gamma_0^{-1}[\ketbra{\psi}] = \begin{bmatrix}
			2|a_0|^2-|a_1|^2 & 3a_0a_1^*\\
			3a_0^*a_1 & 2|a_1|^2-|a_0|^2
		\end{bmatrix}.
	\end{equation*}
	It is known that $\Gamma_0$ has an inverse as it is a linear mapping. Applying $\Gamma_0^{-1}$ on $\ket{B_j}$, we have, again by direct calculation, that
	\begin{equation*}
		\Gamma_0^{-1}\left[\ketbra{B_j}\right]=2\ketbra{B_j}-\ketbra{1-{B_j}}.
	\end{equation*}
	
	By (\ref{eq:seed}), taking the expectation of $W_j\ketbra{C_j}$ gives
	\begin{equation*}
		\EE\Big[W_j\ketbra{C_j}\Big] =\frac{2}{3} \ketbra{B_j} - \frac{1}{3} \ketbra{1-B_j} = \frac{1}{3} \Gamma_0^{-1}\left[\ketbra{B_j}\right].
	\end{equation*}
	Since $\ket{V_j}=U_j\ket{C_j}$, we get
	\begin{equation}
		\label{eq:o-2}
		\EE\left[3 W_j\ketbra{V_j}\right] = U_j\Gamma_0^{-1}\left[\ketbra{B_j}\right]U_j^\dagger  = \Gamma_0^{-1}\left[U_j\ketbra{B_j}U_j^\dagger\right].
	\end{equation}
	Since $(U_j, B_j)$'s are independent for different $j \in [n]$, 
	\begin{eqnarray}
		\EE\left[3^n W \ketbra{\tilde{\varphi}_i}\right] &=& \EE\left[ \bigotimes_{j=1}^n \left(3 W_j\ketbra{v_j}\right)\right] \\
		&= & \EE\left[ \bigotimes_{j=1}^n \Gamma_0^{-1}[U_j\ketbra{B_j}U_j^\dagger]\right] \nonumber \\
		&=& \bigotimes_{j=1}^n \EE\left[\Gamma_0^{-1}\left[U_j\ketbra{B_j}U_j^\dagger\right]\right] \nonumber
		\\
		&=& \ketbra{\phi}. \label{eq:o-3}
	\end{eqnarray}
	where the last equality follows form \cite[Lemma 6]{HNX+23}.   
\end{proof}

The next lemma shows that $S$ is an unbiased estimator of the quantity $\ev{M}{\phi}$.
\begin{lemma}
	\label{lem:unbiased-S}
	$\E[S] = \ev{M}{\phi}$ .
\end{lemma}

\begin{proof}
	Without loss of generality, assume that $q_\ell = \ell$ for all $\ell \in [k]$. We have
	\begin{eqnarray}
		\EE[S] &=& 3^k \EE\left[X \prod_{j \in [k]} W_j \right] \nonumber \\
		&=&  3^k \EE\left[\ev{M}{\tilde{\phi}} \prod_{j \in [k]} W_j\right] \nonumber \\
		&=&  3^k \EE\left[\tr\left(M \ketbra{\tilde{\phi}} \right)  \prod_{j \in [k]} W_j\right] \nonumber \\
		&=&\tr\left\{M  \EE\left[\bigotimes_{j \in [k]} \left(3 W_j \ketbra{V_j}\right) \bigotimes \ketbra{0^{n-k}}\right] \right\}. \quad\quad \label{eq:q-3}
	\end{eqnarray}
	Since $M$ only depends on the first $k$ qubits, the expectation in (\ref{eq:q-3}) does not change if $\ketbra{0^{n-k}}$ is replaced by the following state 
	\begin{equation*}
		\bigotimes_{j=k+1}^n \EE\left[\Gamma_0^{-1}\left[U_j \ketbra{b_j}U_j^\dagger\right]\right].
	\end{equation*}
	Hence, we can write $\EE[S]$ as
	\begin{eqnarray*}
		&&\tr\left\{M   \EE\left[\bigotimes_{j=1}^k \left(3W_{i,j} \ketbra{V_{i,j}}\right)\right]    \bigotimes_{j=k+1}^n \EE\left[\Gamma_0^{-1}\left[U_j \ketbra{B_j}U_j^\dagger\right]\right]  \right\}       \\
		&\stackrel{}{=}&  \tr\left\{M \bigotimes_{j=1}^n \EE\left[\Gamma_0^{-1}\left[U_j \ketbra{B_j}U_j^\dagger\right]\right]  \right\}\\
		&\stackrel{}{=}& \tr{M \ketbra{\phi}} = \ev{M}{\phi},
	\end{eqnarray*}
where the second equality follows from (\ref{eq:o-2}), and the third equality follows from (\ref{eq:o-3}).
\end{proof}

The correctness part of Theorem~\ref{thm:QSS} follows immediately from Lemma~\ref{lem:unbiased-S}, Hoeffding's inequality (Lemma \ref{lem:hoeffding}), and the fact that for all $i \in [N]$, we have $\abs{S_i} \le 3^k \norm{M}_\infty$.

\paragraph{Running time}  We now analyze the running time of the query estimation procedure.  First, the preparation of each quantum state $\ket{v'_{i,j}}\ (i \in [N], j \in [n])$ takes quantum time $O(1)$.  The construction of each shadow sample $\ket{\tilde{\phi}_i}$ (Eq.~(\ref{eq:p-11})) takes $O(k)$ quantum time; note that we do not actually need to prepare those $\ket{v'_{i,j}}$'s with $j \not\in Q$, since the $k$-local observable does not depend on those qubits. 
The quantum time for measuring each $\ket{\tilde{\phi}_i}$ with a $k$-local observable $M$ is $O\left(poly(k)\right)$, as we have assumed that $M$ has a $\text{poly}(k)$ gate complexity.   The computation of each $S_i$ (Eq.~(\ref{eq:p-2})) can be done in $O(n)$ classical time, and 
 that of $T = \frac{1}{N}\sum_i S_i$ can be bounded by $O(N)$ classical time.  Summing up everything, the total running time can be bounded by  $O\left(poly(k) N\right)$ quantum time plus $O(k N)$ classical time; the classical time can be ignored if we assume that a unit quantum time is at least a unit classical time.

%% file: paper.bbl
\begin{thebibliography}{10}

\bibitem{Aaronson07}
Scott Aaronson.
\newblock The learnability of quantum states.
\newblock {\em Proceedings of the Royal Society A: Mathematical, Physical and
  Engineering Sciences}, 463(2088):3089--3114, 2007.

\bibitem{Aaronson18}
Scott Aaronson.
\newblock Shadow tomography of quantum states.
\newblock In Ilias Diakonikolas, David Kempe, and Monika Henzinger, editors,
  {\em STOC}, pages 325--338. {ACM}, 2018.

\bibitem{Aaronson2004}
Scott Aaronson and Daniel Gottesman.
\newblock Improved simulation of stabilizer circuits.
\newblock {\em Physical Review A}, 70(5):052328, nov 2004.
\newblock \href {http://dx.doi.org/10.1103/physreva.70.052328}
  {\path{doi:10.1103/physreva.70.052328}}.

\bibitem{ACNN11}
Alexandr Andoni, Moses Charikar, Ofer Neiman, and Huy~L. Nguyen.
\newblock Near linear lower bound for dimension reduction in {L1}.
\newblock In Rafail Ostrovsky, editor, {\em FOCS}, pages 315--323. {IEEE}
  Computer Society, 2011.

\bibitem{AI06}
Alexandr Andoni and Piotr Indyk.
\newblock Near-optimal hashing algorithms for approximate nearest neighbor in
  high dimensions.
\newblock In {\em FOCS}, pages 459--468. {IEEE} Computer Society, 2006.

\bibitem{BO21}
Costin Badescu and Ryan O'Donnell.
\newblock Improved quantum data analysis.
\newblock In Samir Khuller and Virginia~Vassilevska Williams, editors, {\em
  STOC}, pages 1398--1411. {ACM}, 2021.

\bibitem{BOW17}
Costin Badescu, Ryan O'Donnell, and John Wright.
\newblock Quantum state certification.
\newblock {\em CoRR}, abs/1708.06002, 2017.

\bibitem{BCW01}
Harry Buhrman, Richard Cleve, John Watrous, and Ronald De~Wolf.
\newblock Quantum fingerprinting.
\newblock {\em Physical Review Letters}, 87(16):167902, 2001.

\bibitem{CGG+23}
Umut {\c{C}}alikyilmaz, Sven Groppe, Jinghua Groppe, Tobias Winker, Stefan
  Prestel, Farida Shagieva, Daanish Arya, Florian Preis, and Le~Gruenwald.
\newblock Opportunities for quantum acceleration of databases: Optimization of
  queries and transaction schedules.
\newblock {\em Proc. {VLDB} Endow.}, 16(9):2344--2353, 2023.

\bibitem{Canonne20}
Clément~L. Canonne.
\newblock A short note on learning discrete distributions, 2020.
\newblock \href {http://arxiv.org/abs/2002.11457} {\path{arXiv:2002.11457}}.

\bibitem{CNY23}
Thomas Chen, Shivam Nadimpalli, and Henry Yuen.
\newblock Testing and learning quantum juntas nearly optimally.
\newblock In Nikhil Bansal and Viswanath Nagarajan, editors, {\em SODA}, pages
  1163--1185.

\bibitem{CL21}
Kai{-}Min Chung and Han{-}Hsuan Lin.
\newblock Sample efficient algorithms for learning quantum channels in {PAC}
  model and the approximate state discrimination problem.
\newblock In Min{-}Hsiu Hsieh, editor, {\em TQC}, volume 197 of {\em LIPIcs},
  pages 3:1--3:22. Schloss Dagstuhl - Leibniz-Zentrum f{\"{u}}r Informatik,
  2021.

\bibitem{Cockshott97}
Paul Cockshott.
\newblock Quantum relational databases, 1997.
\newblock \href {http://arxiv.org/abs/quant-ph/9712025}
  {\path{arXiv:quant-ph/9712025}}.

\bibitem{Dankert2009}
Christoph Dankert, Richard Cleve, Joseph Emerson, and Etera Livine.
\newblock Exact and approximate unitary 2-designs and their application to
  fidelity estimation.
\newblock {\em Physical Review A}, 80(1):012304, July 2009.
\newblock \href {http://dx.doi.org/10.1103/physreva.80.012304}
  {\path{doi:10.1103/physreva.80.012304}}.

\bibitem{DII+04}
Mayur Datar, Nicole Immorlica, Piotr Indyk, and Vahab~S. Mirrokni.
\newblock Locality-sensitive hashing scheme based on p-stable distributions.
\newblock In Jack Snoeyink and Jean{-}Daniel Boissonnat, editors, {\em SOCG},
  pages 253--262. {ACM}, 2004.

\bibitem{DiVincenzo2002}
D.P. DiVincenzo, D.W. Leung, and B.M. Terhal.
\newblock Quantum data hiding.
\newblock {\em IEEE Transactions on Information Theory}, 48(3):580--598, March
  2002.
\newblock \href {http://dx.doi.org/10.1109/18.985948}
  {\path{doi:10.1109/18.985948}}.

\bibitem{Farhi2014}
Edward Farhi, Jeffrey Goldstone, and Sam Gutmann.
\newblock A quantum approximate optimization algorithm, 2014.
\newblock \href {http://arxiv.org/abs/1411.4028} {\path{arXiv:1411.4028}},
  \href {http://dx.doi.org/10.48550/ARXIV.1411.4028}
  {\path{doi:10.48550/ARXIV.1411.4028}}.

\bibitem{Farhi2018}
Edward Farhi and Hartmut Neven.
\newblock Classification with quantum neural networks on near term processors.
\newblock February 2018.
\newblock \href {http://arxiv.org/abs/1802.06002} {\path{arXiv:1802.06002}}.

\bibitem{FBK21}
Daniel~Stilck Fran{\c{c}}a, Fernando G. S.~L. Brand{\~{a}}o, and Richard Kueng.
\newblock Fast and robust quantum state tomography from few basis measurements.
\newblock In Min{-}Hsiu Hsieh, editor, {\em 16th Conference on the Theory of
  Quantum Computation, Communication and Cryptography, {TQC} 2021, July 5-8,
  2021, Virtual Conference}, volume 197 of {\em LIPIcs}, pages 7:1--7:13.
  Schloss Dagstuhl - Leibniz-Zentrum f{\"{u}}r Informatik, 2021.

\bibitem{Garg2020}
Siddhant Garg and Goutham Ramakrishnan.
\newblock Advances in quantum deep learning: An overview.
\newblock {\em arXiv:2005.04316}, May 2020.
\newblock \href {http://arxiv.org/abs/2005.04316} {\path{arXiv:2005.04316}}.

\bibitem{GA22}
Weiyuan Gong and Scott Aaronson.
\newblock Learning distributions over quantum measurement outcomes.
\newblock {\em CoRR}, abs/2209.03007, 2022.

\bibitem{Grover96}
Lov~K. Grover.
\newblock A fast quantum mechanical algorithm for database search.
\newblock In Gary~L. Miller, editor, {\em STOC}, pages 212--219. {ACM}, 1996.

\bibitem{HHJ+16}
Jeongwan Haah, Aram~W. Harrow, Zheng{-}Feng Ji, Xiaodi Wu, and Nengkun Yu.
\newblock Sample-optimal tomography of quantum states.
\newblock In Daniel Wichs and Yishay Mansour, editors, {\em STOC}, pages
  913--925. {ACM}, 2016.

\bibitem{Haah2017}
Jeongwan Haah, Aram~W. Harrow, Zhengfeng Ji, Xiaodi Wu, and Nengkun Yu.
\newblock Sample-optimal tomography of quantum states.
\newblock {\em {IEEE} Transactions on Information Theory}, pages 1--1, 2017.
\newblock \href {http://dx.doi.org/10.1109/tit.2017.2719044}
  {\path{doi:10.1109/tit.2017.2719044}}.

\bibitem{HHF24}
Rihan Hai, Shih{-}Han Hung, and Sebastian Feld.
\newblock Quantum data management: From theory to opportunities.
\newblock In {\em ICDE}, pages 5376--5381. {IEEE}, 2024.
\newblock URL: \url{https://doi.org/10.1109/ICDE60146.2024.00410}, \href
  {http://dx.doi.org/10.1109/ICDE60146.2024.00410}
  {\path{doi:10.1109/ICDE60146.2024.00410}}.

\bibitem{HHL09}
Aram~W Harrow, Avinatan Hassidim, and Seth Lloyd.
\newblock Quantum algorithm for linear systems of equations.
\newblock {\em Physical review letters}, 103(15):150502, 2009.

\bibitem{HLM17}
Aram~W. Harrow, Cedric~Yen{-}Yu Lin, and Ashley Montanaro.
\newblock Sequential measurements, disturbance and property testing.
\newblock In Philip~N. Klein, editor, {\em SODA}, pages 1598--1611. {SIAM},
  2017.

\bibitem{Harrow2021}
Aram~W. Harrow and John~C. Napp.
\newblock Low-depth gradient measurements can improve convergence in
  variational hybrid quantum-classical algorithms.
\newblock {\em Physical Review Letters}, 126(14):140502, apr 2021.
\newblock \href {http://dx.doi.org/10.1103/physrevlett.126.140502}
  {\path{doi:10.1103/physrevlett.126.140502}}.

\bibitem{HW12}
Aram~W. Harrow and Andreas~J. Winter.
\newblock How many copies are needed for state discrimination?
\newblock {\em {IEEE} Trans. Inf. Theory}, 58(1):1--2, 2012.

\bibitem{HSW08}
Patrick Hayden, Peter~W. Shor, and Andreas Winter.
\newblock Random quantum codes from gaussian ensembles and an uncertainty
  relation.
\newblock {\em Open Systems \&\ Information Dynamics}, 15(01):71--89, mar 2008.

\bibitem{HeidariAAAI2022}
Mohsen Heidari, Ananth~Y. Grama, and Wojciech Szpankowski.
\newblock Toward physically realizable quantum neural networks.
\newblock {\em Association for the Advancement of Articial Intelligence
  (AAAI)}, 2022.

\bibitem{HNX+23}
Mohsen Heidari, Mobasshir~A Naved, Zahra Honjani, Wenbo Xie, Arjun~Jacob Grama,
  and Wojciech Szpankowski.
\newblock Quantum shadow gradient descent for variational quantum algorithms.
\newblock 2024.
\newblock URL: \url{https://arxiv.org/abs/2310.06935}, \href
  {http://arxiv.org/abs/2310.06935} {\path{arXiv:2310.06935}}.

\bibitem{Heidari2021}
Mohsen Heidari, Arun Padakandla, and Wojciech Szpankowski.
\newblock A theoretical framework for learning from quantum data.
\newblock In {\em {IEEE} International Symposium on Information Theory
  ({ISIT})}, 2021.

\bibitem{HP06}
Fumio Hiai and Denes Petz.
\newblock {\em The Semicircle Law, Free Random Variables and Entropy
  (Mathematical Surveys \&\ Monographs)}.
\newblock American Mathematical Society, 2006.

\bibitem{Holevo13}
Alexander~S. Holevo.
\newblock {\em Quantum Systems, Channels, Information}.
\newblock De Gruyter, Berlin, Boston, 2013.
\newblock URL: \url{https://doi.org/10.1515/9783110273403} [cited 2023-08-27],
  \href {http://dx.doi.org/doi:10.1515/9783110273403}
  {\path{doi:doi:10.1515/9783110273403}}.

\bibitem{Holevo78}
Alexander~Semenovich Holevo.
\newblock On asymptotically optimal hypotheses testing in quantum statistics.
\newblock {\em Teoriya Veroyatnostei i ee Primeneniya}, 23(2):429--432, 1978.

\bibitem{HKP20}
Hsin-Yuan Huang, Richard Kueng, and John Preskill.
\newblock Predicting many properties of a quantum system from very few
  measurements.
\newblock {\em Nature Physics 16, 1050--1057 (2020)}, February 2020.

\bibitem{HPW24}
Noah Huffman, Dmitri Pavlichin, and Tsachy Weissman.
\newblock Lossy compression for schr\"odinger-style quantum simulations, 2024.
\newblock URL: \url{https://arxiv.org/abs/2401.11088}, \href
  {http://arxiv.org/abs/2401.11088} {\path{arXiv:2401.11088}}.

\bibitem{IM98}
Piotr Indyk and Rajeev Motwani.
\newblock Approximate nearest neighbors: Towards removing the curse of
  dimensionality.
\newblock In Jeffrey~Scott Vitter, editor, {\em STOC}, pages 604--613. {ACM},
  1998.

\bibitem{JZK+03}
E.~Joos, H.~D. Zeh, C.~Kiefer, D.~J.~W. Giulini, J.~Kupsch, and I.~O.
  Stamatescu.
\newblock {\em Decoherence and the Appearance of a Classical World in Quantum
  Theory}.
\newblock Springer, 2003.

\bibitem{Jordan2012}
Stephen~P. Jordan, Keith S.~M. Lee, and John Preskill.
\newblock Quantum algorithms for quantum field theories.
\newblock {\em Science}, 336(6085):1130--1133, jun 2012.
\newblock \href {http://dx.doi.org/10.1126/science.1217069}
  {\path{doi:10.1126/science.1217069}}.

\bibitem{KKR06}
Julia Kempe, Alexei~Y. Kitaev, and Oded Regev.
\newblock The complexity of the local hamiltonian problem.
\newblock {\em {SIAM} J. Comput.}, 35(5):1070--1097, 2006.

\bibitem{Kivlichan2020}
Ian~D. Kivlichan, Craig Gidney, Dominic~W. Berry, Nathan Wiebe, Jarrod McClean,
  Wei Sun, Zhang Jiang, Nicholas Rubin, Austin Fowler, Al{\'{a}}n Aspuru-Guzik,
  Hartmut Neven, and Ryan Babbush.
\newblock Improved fault-tolerant quantum simulation of condensed-phase
  correlated electrons via trotterization.
\newblock {\em Quantum}, 4:296, jul 2020.
\newblock \href {http://dx.doi.org/10.22331/q-2020-07-16-296}
  {\path{doi:10.22331/q-2020-07-16-296}}.

\bibitem{Liu07}
Yang Liu and Gui~Lu Long.
\newblock Deleting a marked item from an unsorted database with a single query,
  2007.
\newblock \href {http://arxiv.org/abs/0710.3301} {\path{arXiv:0710.3301}}.

\bibitem{Massoli2021}
Fabio~Valerio Massoli, Lucia Vadicamo, Giuseppe Amato, and Fabrizio Falchi.
\newblock A leap among entanglement and neural networks: A quantum survey.
\newblock {\em arXiv:2107.03313}, July 2021.
\newblock \href {http://arxiv.org/abs/2107.03313} {\path{arXiv:2107.03313}}.

\bibitem{NC10}
Isaac L.~Chuang Michael A.~Nielsen.
\newblock {\em Quantum Computation and Quantum Information}.
\newblock Cambridge University Pr., December 2010.

\bibitem{MichaelA.Nielsen2010}
Isaac L.~Chuang Michael A.~Nielsen.
\newblock {\em Quantum Computation and Quantum Information}.
\newblock Cambridge University Pr., December 2010.
\newblock URL:
  \url{https://www.ebook.de/de/product/13055864/michael_a_nielsen_isaac_l_chuang_quantum_computation_and_quantum_information.html}.

\bibitem{Mitarai2018}
K.~Mitarai, M.~Negoro, M.~Kitagawa, and K.~Fujii.
\newblock Quantum circuit learning.
\newblock {\em Physical Review A}, 98(3):032309, sep 2018.
\newblock \href {http://dx.doi.org/10.1103/physreva.98.032309}
  {\path{doi:10.1103/physreva.98.032309}}.

\bibitem{OW16}
Ryan O'Donnell and John Wright.
\newblock Efficient quantum tomography.
\newblock In Daniel Wichs and Yishay Mansour, editors, {\em STOC}, pages
  899--912. {ACM}, 2016.

\bibitem{PAB+23}
Alexey Pyrkov, Alex Aliper, Dmitry Bezrukov, Yen-Chu Lin, Daniil Polykovskiy,
  Petrina Kamya, Feng Ren, and Alex Zhavoronkov.
\newblock Quantum computing for near-term applications in generative chemistry
  and drug discovery.
\newblock {\em Drug Discovery Today}, page 103675, 2023.

\bibitem{SC95}
Jun~John Sakurai and Eugene~D Commins.
\newblock Modern quantum mechanics, revised edition, 1995.

\bibitem{Schuld2014}
Maria Schuld, Ilya Sinayskiy, and Francesco Petruccione.
\newblock The quest for a quantum neural network.
\newblock {\em Quantum Information Processing}, 13(11):2567–2586, Aug 2014.
\newblock \href {http://dx.doi.org/10.1007/s11128-014-0809-8}
  {\path{doi:10.1007/s11128-014-0809-8}}.

\bibitem{Sen06}
Pranab Sen.
\newblock Random measurement bases, quantum state distinction and applications
  to the hidden subgroup problem.
\newblock In {\em CCC}, pages 274--287. {IEEE} Computer Society, 2006.

\bibitem{Shor97}
Peter~W. Shor.
\newblock Polynomial-time algorithms for prime factorization and discrete
  logarithms on a quantum computer.
\newblock {\em {SIAM} J. Comput.}, 26(5):1484--1509, 1997.

\bibitem{SKPK19}
Adam Smith, MS~Kim, Frank Pollmann, and Johannes Knolle.
\newblock Simulating quantum many-body dynamics on a current digital quantum
  computer.
\newblock {\em npj Quantum Information}, 5(1):106, 2019.

\bibitem{STY+23}
Xiaoming Sun, Guojing Tian, Shuai Yang, Pei Yuan, and Shengyu Zhang.
\newblock Asymptotically optimal circuit depth for quantum state preparation
  and general unitary synthesis, 2023.
\newblock \href {http://arxiv.org/abs/2108.06150} {\path{arXiv:2108.06150}}.

\bibitem{TK16}
Immanuel Trummer and Christoph Koch.
\newblock Multiple query optimization on the d-wave 2x adiabatic quantum
  computer.
\newblock {\em Proc. {VLDB} Endow.}, 9(9):648--659, 2016.
\newblock URL: \url{http://www.vldb.org/pvldb/vol9/p648-trummer.pdf}, \href
  {http://dx.doi.org/10.14778/2947618.2947621}
  {\path{doi:10.14778/2947618.2947621}}.

\bibitem{Whitfield2011}
James~D. Whitfield, Jacob Biamonte, and Al{\'{a}}n Aspuru-Guzik.
\newblock Simulation of electronic structure hamiltonians using quantum
  computers.
\newblock {\em Molecular Physics}, 109(5):735--750, mar 2011.
\newblock \href {http://dx.doi.org/10.1080/00268976.2011.552441}
  {\path{doi:10.1080/00268976.2011.552441}}.

\bibitem{WGU+23}
Tobias Winker, Sven Groppe, Valter Uotila, Zhengtong Yan, Jiaheng Lu, Maja
  Franz, and Wolfgang Mauerer.
\newblock Quantum machine learning: Foundation, new techniques, and
  opportunities for database research.
\newblock In Sudipto Das, Ippokratis Pandis, K.~Sel{\c{c}}uk Candan, and Sihem
  Amer{-}Yahia, editors, {\em SIGMOD}, pages 45--52. {ACM}, 2023.

\bibitem{WDD+19}
Xin{-}Chuan Wu, Sheng Di, Emma~Maitreyee Dasgupta, Franck Cappello, Hal Finkel,
  Yuri Alexeev, and Frederic~T. Chong.
\newblock Full-state quantum circuit simulation by using data compression.
\newblock In Michela Taufer, Pavan Balaji, and Antonio~J. Pe{\~{n}}a, editors,
  {\em SC}, pages 80:1--80:24. {ACM}, 2019.

\bibitem{Younes07}
Ahmed Younes.
\newblock Database manipulation on quantum computers, 2007.
\newblock \href {http://arxiv.org/abs/0705.4303} {\path{arXiv:0705.4303}}.

\end{thebibliography}
